\documentclass[journal=jacsat,manuscript=article,a4paper,12pt]{achemso}

\usepackage[version=3]{mhchem}
\usepackage[normalem]{ulem}
\usepackage[colorlinks]{hyperref}

\newcommand{\de}{{\rm d}}
\newcommand{\br}{{\boldsymbol{r}}}
\newcommand{\beq}{\begin{equation}}
\newcommand{\eeq}{\end{equation}}

\newcommand\blfootnote[1]{%
	\begingroup%
	\renewcommand\thefootnote{}\footnote{#1}%
	\addtocounter{footnote}{-1}%
	\endgroup%
}

\usepackage[dvipsnames]{xcolor}
\usepackage{amssymb}

\title{\vspace{-.4 cm}
Quantum-classical dynamical\\
density functional theory for solvation dynamics\vspace{-.15cm}}

\author{\normalsize Paul Bergold}
\email{paul.bergold@ug.edu.pl}
\affiliation{\footnotesize Instytut Matematyki, Uniwersytet Gda\'nski, Gda\'nsk, Poland}

\author{Irene Burghardt}
\email{burghardt@chemie.uni-frankfurt.de}
\affiliation{\footnotesize Institute of Physical and Theoretical Chemistry, Goethe University Frankfurt, Frankfurt,
Germany}

\author{Francesco Di Maiolo}
\email{francesco.dimaiolo@unipr.it}
\affiliation{\footnotesize Department of Chemistry, Life Science and Environmental Sustainability, Universit\`a di Parma, Italy}

\author{Cesare Tronci\vspace{-.25cm}}
\email{c.tronci@surrey.ac.uk}
\affiliation{\footnotesize School of Mathematics and Physics, University of Surrey, Guildford, United Kingdom}
\alsoaffiliation{\footnotesize School of Physical and Mathematical Sciences, Nanyang Technological University, Singapore\vspace{-.15cm}}

\abbreviations{QC, DDFT, QCLE, DCM, DMSO, FWHM}
\SectionNumbersOn
\setkeys{acs}{maxauthors=10,etalmode=truncate}

\begin{document}
\begin{abstract}\vspace{-.8cm}
	\noindent We\blfootnote{All authors contributed equally to this work.} provide a unified modeling framework for thermalization and non-equilibrium spectroscopy in overdamped polar environments.
	This framework extends dynamical density functional theory to describe the coupled quantum-classical evolution of a molecular electronic subsystem and an overdamped polar solvent represented by a reaction-field coordinate.
	The nonadiabatic dynamics is formulated in terms of an operator-valued distribution obeying a quantum-classical Ehrenfest--Smoluchowski equation which exhibits nonlinear classical diffusion and non-increasing free energy, thereby retaining detailed balance.
	In contrast, a quantum-classical Liouville--Smoluchowski model with quantum-classical linear diffusion develops unstable long-time behavior and fails to reach a stationary thermal state.
	The method is applied to a two-level model of a polar dye in three different solvents: chloroform, dichloromethane, and dimethyl sulfoxide, each differing in polarity and dielectric relaxation time.
	Absorption and time-resolved emission spectra are obtained from dipole-dipole correlation functions, capturing the solvent-induced spectral broadening, the dynamic Stokes shift, and the time evolution of the emission bandwidth.
	The calculated solvent spectral response agrees well with the experimental relaxation trends of the three solvents.
\end{abstract}
%

\section{Introduction}\label{sec:intro}
\subsection{Solute-solvent interaction in solvation dynamics}
Solvent environments can considerably reshape the energetics, dynamics, and optical response of molecular systems in solution.
These effects arise because the solute and solvent continuously influence one another: the molecular charge distribution polarizes the surrounding medium, while fluctuations and relaxation of the solvent modify the effective energetic landscape experienced by the solute.
This interplay is relevant both at equilibrium, where the solvent samples a thermal distribution of configurations around the solute, and under non-equilibrium conditions, where the environment relaxes in response to a change in the molecular state.\cite{Maroncelli1989,Rosspeintner2013,VenkatramanOrrEwing2021}

For polar solutes, the response of the surrounding medium to the molecular charge distribution can be conveniently described within a dielectric continuum picture.
In this framework, the polarization of the solvent is represented through the reaction field acting on the solute.\cite{Tomasi2005,TerenzianiPainelli2003,Burghardt2006}
Two contributions to this field are usually distinguished: a fast component associated with the electronic polarization of the solvent and a nuclear component related primarily to molecular reorientation.
The latter can itself display distinct dynamical regimes.
At very short times, the solvent response may contain an inertial component associated with collective librational motion, whereas at longer times it evolves toward predominantly diffusive relaxation.\cite{Maroncelli1989,Rosspeintner2013,VenkatramanOrrEwing2021,Collini2018}
The present work focuses on this latter regime, where the nuclear solvent response can be described in terms of overdamped dynamics.
Dipolar donor-acceptor dyes are particularly sensitive to solvent polarization because their ground and excited electronic states can differ markedly in charge distribution and permanent dipole moment, leading to pronounced solvatochromic effects.\cite{Reichardt1994,TerenzianiPainelli2003,Bardi2023}

The same solute-solvent coupling governs both relaxation toward thermal equilibrium and photoinduced dynamics.
If the solvent distribution is initially out of equilibrium with a given electronic state, dissipative dynamics drives the coupled system toward the corresponding stationary state.
Following photoexcitation, the non-equilibrium character of the problem becomes especially apparent: the electronic charge distribution changes before the nuclear solvent polarization can reorganize, so that the solvent initially retains the configuration associated with the ground state of the dye and subsequently relaxes in response to the excited state charge distribution.\cite{Maroncelli1989,Rosspeintner2013,VenkatramanOrrEwing2021}
This reorganization changes the stabilization of the electronic states and is directly reflected in time-dependent optical observables.
In particular, solvent relaxation produces the dynamic Stokes shift, while thermal fluctuations of the reaction field contribute to the inhomogeneous broadening of optical spectra.\cite{Maroncelli1989,Horng1995,Bardi2023}

Recent experiments have provided increasingly direct access to this coupled solute-solvent dynamics.
Ultrafast optical spectroscopy has resolved the initial stages of solvent relaxation, while time-resolved X-ray scattering has revealed structural rearrangements of the surrounding solvent driven by photoinduced charge redistribution and electron transfer.\cite{Collini2018,Biasin2021,Winte2025,Mitterer2026}
These observations reinforce the view of the solvent as a dynamical component of the relaxation process and motivate theoretical descriptions able to treat the electronic dynamics of the solute together with the dissipative response of the environment.

To describe this dissipative regime, the collective dielectric response can be mapped onto an effective reaction-field coordinate.
Within the linear-response regime, the corresponding free-energy surfaces are quadratic functions of this coordinate, while the solvent state is described by a probability distribution over reaction-field values.
The dynamics of this distribution is conventionally described by a Smoluchowski equation.\cite{CherniakMukamel1996,TerenzianiPainelli2003,DiMaioloPainelli2020,Bardi2023}
Its characteristic timescale is related to the longitudinal relaxation time of the solvent, $\tau_L$.
This effective coordinate framework provides a compact description of the dissipative solvent response and connects solvent relaxation with the time-dependent evolution of optical spectra.\cite{TerenzianiPainelli2003,terenziani2015,DiMaioloPainelli2020}
The situation becomes more subtle when the electronic subsystem is allowed to evolve dynamically rather than being represented only through state-specific free-energy surfaces.
The instantaneous reaction field modifies the electronic Hamiltonian, while the electronic state determines, in turn, the forces driving the solvent coordinate.
Here, we develop a mixed quantum-classical (MQC) description that retains the quantum evolution of the electronic subsystem together with the overdamped, dissipative relaxation of the classical environment.

\subsection{A quantum-classical Liouville--Smoluchowski model}
Following previous proposals \cite{BoEtAl11,Burghardt2006,HuEtAl12}, our approach is motivated by the free-energy-driven structure of dynamical density functional theory (DDFT).\cite{MarconiTarazona1999,ArcherEvans2004,teVrugt2020} 
In its standard formulation, DDFT describes the overdamped evolution of a classical spatial particle density through thermodynamic forces derived from a free-energy functional, and is closely connected to the underlying Smoluchowski dynamics.\cite{ArcherEvans2004}
The classical variable considered here, however, has a different meaning: it is not the spatial particle density of the solvent, but a probability distribution over an effective reaction-field coordinate.
Despite this difference from the standard approach, the thermodynamic free energy $F=E-TS$ remains a natural quantity for describing the solvent dynamics.
As usual, here $E$ denotes the total energy, while $S$ is the entropy and $T$ the reference equilibrium temperature.

In this work, we introduce a positive-semidefinite operator-valued distribution $\widehat P(\br,t)$ over the classical reaction-field coordinates $\br$, such that $\operatorname{Tr}\int\widehat P(\br,t)\,\de\br=1$.
Its matrix structure resolves electronic populations and coherences with respect to the solvent coordinate, while tracing over the electronic degrees of freedom yields the corresponding classical reaction-field distribution.
The MQC dynamics of $\widehat P(\br,t)$ therefore provides a description of the coupled evolution of the quantum electronic subsystem and the classical solvent coordinate.

An early example of such an MQC formulation is the quantum-classical Redfield--Smo\-lu\-chow\-ski equation introduced in Ref.~\citenum{DiMaioloPainelli2020}.
While the original expression includes an additional Redfield-type quantum thermostat, here we omit it to focus on the coupled solute-solvent dynamics.
In the ground state, this corresponds to assuming that the solute is already at vibrational equilibrium.
In the excited state, we exploit the separation between the characteristic timescales of intramolecular vibrational relaxation and solvent reorganization, so that vibrational contributions to the time evolution of the spectra can be approximately disentangled from the solvation dynamics.\cite{Horng1995}
We refer to the resulting modified equation as the quantum-classical Liouville--Smoluchowski (QCLS) equation.
For an arbitrary energy functional $E=E(\widehat{P})$, this reads
\begin{equation}\label{eq:qcls}
	i\hbar\frac{\partial\widehat{P}}{\partial t}
	=i\hbar\mu\operatorname{div}\left(\widehat{P}\left\langle\nabla\frac{\delta E}{\delta\widehat{P}}\right\rangle+k_BT\nabla\widehat{P}\right)+\left[\frac{\delta E}{\delta\widehat{P}},\widehat{P}\right].
\end{equation}
Here, $\delta E/\delta\widehat{P}$ denotes the functional derivative of the energy functional with respect to the operator-valued distribution $\widehat{P}$, while spatial differential operators act on the reaction-field coordinates $\br$.
Moreover, $\mu$ and $k_B$ are the solvent mobility and the Boltzmann constant, respectively.
We have also introduced the notation
\begin{equation}\label{eq:local-expectation}
	\langle\widehat{A}\rangle
	=\frac{\operatorname{Tr}(\widehat{A}\widehat{P})}{\operatorname{Tr}\widehat{P}}, 
\end{equation}
where the trace is taken over the electronic subsystem Hilbert space.
Then, the energy functional $E=\operatorname{Tr}{\int}\widehat{H}\widehat{P}\,\de\br$ leads to the averaged Hellmann--Feynman forces $\mathbf{F}=-\langle\nabla\widehat{H}\rangle$ characteristic of the Ehrenfest approximation.
This averaging procedure was adopted in Ref.~\citenum{DiMaioloPainelli2020} to avoid the negative probability values that may arise from the original derivation based on the quantum-classical Liouville equation \cite{Kapral2006} and its collisional Fokker--Planck variant.\cite{Kapral2001}
Besides the absence of the Redfield term, Eq.~\eqref{eq:qcls} differs from the Redfield--Smoluchowski model of Ref.~\citenum{DiMaioloPainelli2020} also in the form of the transport term.
Specifically, the term $\langle\nabla{\delta E}/{\delta\widehat{P}}\rangle\cdot\nabla\widehat{P}$ used in Ref.~\citenum{DiMaioloPainelli2020} is replaced here by $\operatorname{div}(\widehat{P}\langle\nabla{\delta E}/{\delta\widehat{P}}\rangle)$.
This form more naturally reflects the transport of a probability distribution and is consistent with the continuity structure underlying standard DDFT.

The formalism is applied to a dipolar dye in solution, described in terms of its relevant electronic degrees of freedom and their coupling to the classical reaction field.
We consider both the relaxation of the coupled dye-solvent system toward thermal equilibrium in the electronic ground state and its non-equilibrium evolution following preparation in the higher-energy electronic state of the solute.
Equilibrium absorption and time-resolved emission spectra are calculated from dipole-dipole correlation functions.
Within the same dynamical framework, this allows us to describe solvent-induced inhomogeneous broadening of optical spectra and the dynamic Stokes shift, and to compare the resulting solvent spectral response with experiment.

The paper is organized as follows.
Section~\ref{sec:theory} presents the quantum-classical dynamical density functional framework, and Section~\ref{sec:num-implementation} describes its numerical implementation.
Section~\ref{sec:results} introduces the solute-solvent coupled model and discusses ground-state thermalization and the fine structure of the solvent dynamics.
Section~\ref{sec:spectra} presents the absorption and time-resolved emission spectra and compares the solvent spectral response with experiment.
Conclusions are drawn in Section~\ref{sec:conclusions}.

\section{Overdamped quantum-classical dynamics}\label{sec:theory}
\subsection{Thermodynamic consistency}
While the transport term in the overdamped QCLS model in Eq.~\eqref{eq:qcls} retains the characteristic structure of DDFT, the linear diffusion of the operator-valued distribution $\widehat{P}$ introduces a thermodynamic consistency issue.
For general mixed-state evolution, this can be seen as follows: first, we denote by $D=\operatorname{Tr}\widehat{P}$ the classical density and use the von Neumann entropy prescription
\begin{equation}\label{eq:von-neumann-entropy}
	S
	=-k_B\operatorname{Tr}{\int}\widehat{P}\ln\widehat{P}\,\de\br,
\end{equation}
so that the relations $[\widehat{P},{\delta S}/{\delta\widehat{P}}]=k_B[\ln\widehat{P},\widehat P]=0$ and $\langle\nabla{\delta S}/{\delta\widehat{P}}\rangle=-k_BD^{-1}\nabla D$ allow us to rewrite Eq.~\eqref{eq:qcls} in terms of the free energy $F=E-TS$ as
\begin{equation}\label{eq:qcls-free-energy-form}
	i\hbar\frac{\partial\widehat{P}}{\partial t}
	=i\hbar\mu\operatorname{div}\left(\widehat{P}\left\langle\nabla\frac{\delta F}{\delta\widehat{P}}\right\rangle+k_BTD\nabla\frac{\widehat{P}}D\right)+\left[\frac{\delta F}{\delta\widehat{P}},\widehat{P}\right].
\end{equation}
Second, we use the chain-rule relation $\dot{F}=\operatorname{Tr}\int({\delta F}/{\delta\widehat{P}})\partial_t\widehat{P}\,\de\br$ to observe that
\begin{equation}\label{eq:qcls-free-energy-rate}
	\dot{F}
	=-\mu{\int}D\left|\left\langle\nabla\frac{\delta F}{\delta\widehat{P}}\right\rangle\right|^2\de\br+\mu k_BT{\int}\left\langle{\operatorname{div}}\bigg(D\nabla\frac{\delta F}{\delta\widehat{P}}\bigg)\right\rangle\,\de\br.
\end{equation}
We notice that, while the first term in Eq.~\eqref{eq:qcls-free-energy-rate} is manifestly non-positive, the second is not sign-definite in general.
Consequently, as we will verify in our simulations, the QCLS dynamics does not guarantee a monotonic decrease of the free energy.
This situation differs significantly from the free-energy decay property of standard DDFT and signals a possible loss of thermodynamic consistency.
The appearance of consistency issues does not come as a surprise when dealing with MQC dynamics.
For example, negative probabilities and the related violations of the Heisenberg uncertainty principle have been reported in surface hopping approaches\cite{bondarenko} and in the quantum-classical Liouville equation.\cite{Schofield}
Beyond total mean-field factorizations, the Ehrenfest dynamics based on Hellmann--Feynman forces appears as the only conventional model satisfying standard physical consistency requirements.
These include a positive-semidefinite quantum density operator, a positive classical probability distribution, and the correct uncoupled quantum and classical limits.\cite{GBTr2022,GBTr2022b}
Here, we extend its construction to the overdamped regime with the aim of restoring the corresponding thermodynamic properties.
Recent efforts to retain consistency while going beyond the Ehrenfest approximation have been proposed in Refs.~\citenum{BaBeGBTr24,BeMaTr26,GBTr-fluid,GBTr-fluid2} and the associated overdamped variants are left for future work.

Of special interest in general MQC dynamics is also the detailed balance problem,\cite{Mauri,Tully} which is related to the difficulty of characterizing explicit thermal equilibrium distributions in phase space; see Ref.~\citenum{TrMCGB} for a recent advance in this direction.
The quantum-classical DDFT (QC-DDFT) models considered here, however, are written in reaction-field configuration space rather than in phase space, thereby allowing us to retain detailed balance in the treatment below.

\subsection{QC-DDFT and the Ehrenfest--Smoluchowski model}
To retain thermodynamic consistency and resolve the issue discussed above, here we extend the Hellmann--Feynman averaging already employed in Eq.~\eqref{eq:qcls} to the full free-energy gradient.
This leads to the following general QC-DDFT equation
\begin{equation}\label{eq:qc-ddft}
	i\hbar\frac{\partial\widehat{P}}{\partial t}
	=i\hbar\operatorname{div}\left(\widehat{P}\left\langle\mathbb{M}\nabla\frac{\delta F}{\delta\widehat{P}}\right\rangle\right)+\left[\frac{\delta F}{\delta\widehat{P}},\widehat{P}\right],
\end{equation}
where $\mathbb{M}$ is a symmetric positive-definite mobility matrix that may, in general, depend on the full operator-valued distribution $\widehat{P}$.

Eq.~\eqref{eq:qc-ddft} has an underlying bracket structure that combines the standard Poisson bracket of quantum Liouville dynamics with a dissipative bracket of the type previously introduced in Refs.~\citenum{HoPuTr10,HoPuTr08} in the context of porous media flows with anisotropic interactions.
Dissipative brackets have a long history \cite{GBTr-fluid2,Gr84,Ka1984,Mo1984} and provide a general structure-preserving framework for modeling irreversible dynamics, especially in complex fluids and soft matter systems.\cite{BerisEdwards,Ottinger}
In the present case, Eq.~\eqref{eq:qc-ddft} is obtained via the relation $\dot{G}=\{\!\{G,F\}\!\}$, valid for an arbitrary functional $G(\widehat{P})$ and the bracket structure
\begin{equation}\label{eq:qc-ddft-bracket}
	\{\!\{G,F\}\!\}
	=-{\operatorname{Tr}}{\int}\widehat{{P}}\left(\left\langle\nabla\frac{\delta G}{\delta\widehat{{P}}}\right\rangle\cdot\mathbb{M}\left\langle\nabla\frac{\delta F}{\delta\widehat{{P}}}\right\rangle+\frac{i}\hbar\left[\frac{\delta G}{\delta\widehat{{P}}},\frac{\delta F}{\delta\widehat{{P}}}\right]\right)\de\br.
\end{equation}
The first term defines the dissipative contribution to the dynamics, whereas the commutator term generates the reversible quantum evolution.
Since $\mathbb{M}$ is positive definite, the free energy $F=E-TS$ is guaranteed to be non-increasing:
\begin{equation}\label{eq:free-energy-dissipation}
	\dot{F}
	=\{\!\{F,F\}\!\}
	=-{\int}D\left\langle\nabla\frac{\delta F}{\delta\widehat{P}}\right\rangle\cdot\mathbb{M}\left\langle\nabla\frac{\delta F}{\delta\widehat{P}}\right\rangle\de\br
	\le 0.
\end{equation}
Eq.~\eqref{eq:qcls-free-energy-form} and Eq.~\eqref{eq:qc-ddft} can be directly compared by specializing the latter to a constant isotropic mobility $\mathbb{M}=\mu\boldsymbol{1}$.
In addition, a close comparison with the original Eq.~\eqref{eq:qcls} can be obtained by making explicit use of the von Neumann entropy from Eq.~\eqref{eq:von-neumann-entropy} in Eq.~\eqref{eq:qc-ddft}, which yields the quantum-classical Ehrenfest--Smoluchowski equation (QCES):
\begin{equation}\label{eq:qces}
	i\hbar\frac{\partial\widehat{P}}{\partial t}
	=i\hbar\mu\operatorname{div}\bigg(\widehat{P}\left\langle\nabla\frac{\delta E}{\delta\widehat{P}}\right\rangle+k_BT\frac{\widehat{P}}D\nabla D\bigg)+\left[\frac{\delta E}{\delta\widehat{P}},\widehat{P}\right].	
\end{equation}
Compared to Eq.~\eqref{eq:qcls}, the linear diffusion term of the entire operator-valued distribution is replaced by a nonlinear diffusion of the classical density.
This step restores the monotonic free-energy decay established above.
Tracing either Eq.~\eqref{eq:qcls} or Eq.~\eqref{eq:qces} gives $\partial_tD=\mu\operatorname{div}(D\langle\nabla{\delta E}/{\delta\widehat{P}}\rangle)+\mu k_BT\nabla^2D$, thereby retaining linear classical diffusion in both cases.
In the pure-state representation, the wavefunction evolution associated with the QCES is found upon writing $\widehat{P}=|\Psi\rangle\langle\Psi|$, where the ket $|\Psi\rangle=|\Psi(\br,t)\rangle$ denotes a quantum-classical state vector and the bra-ket inner product acts on the quantum components.
Then, $D=\langle\Psi|\Psi\rangle=\|\Psi\|^2$ and Eq.~\eqref{eq:local-expectation} is replaced by $\langle\widehat{A}\rangle=\langle\Psi|\,\widehat{A}\,|\Psi\rangle/D$.
Writing $E={\int}D\langle\widehat{H}\rangle\,\de\br$ takes Eq.~\eqref{eq:qces} into the MQC Schr\"odinger form $i\hbar(\partial_t|\Psi\rangle+\boldsymbol{v}\cdot\nabla|\Psi\rangle+|\Psi\rangle\operatorname{div}\boldsymbol{v}/2)=\widehat{H}|\Psi\rangle$, where ${\boldsymbol{v}}=-\mu\langle\nabla\widehat{H}\rangle-\mu k_BT\nabla D/D$.

Interestingly, Eq.~\eqref{eq:qces} is also obtained by replacing the von Neumann entropy prescription in Eq.~\eqref{eq:von-neumann-entropy} by the classical Gibbs entropy $S_{\mathrm{G}}=-k_B{\int}D\ln D\,\de\br$ associated with the solvent distribution.
For example, in the pure-state representation one has $S=S_{\mathrm{G}}$ and Eq.~\eqref{eq:free-energy-dissipation} again ensures that the corresponding free energy is non-increasing, i.e., $\dot{E}-{T\dot{S}_{\mathrm{G}}\le 0}$.
The fact that Eq.~\eqref{eq:qces} is obtained from Eq.~\eqref{eq:qc-ddft}, regardless of whether one uses $S$ or $S_{\mathrm{G}}$ in the free energy expression, makes the role of the dissipative term in the QCES equation particularly transparent: the irreversible overdamped regime is associated with the classical solvent degrees of freedom, while no quantum entropy enters the diffusion process.
Nevertheless, the quantum and classical subsystems remain dynamically coupled and Eq.~\eqref{eq:free-energy-dissipation} ensures that the solvent dissipation drives the relaxation of the entire solute-solvent system.

A separate discussion is in place about the specific form of the energy functional $E(\widehat{P})$.
Indeed, while in the present treatment the total energy is given by the overall expectation 
\begin{equation}\label{eq:ehrenfest-energy}
	E
	={\int}D\langle\widehat{H}\rangle\,\de\br,
\end{equation}
of the MQC Hamiltonian, extra energy terms can be added to account for further backreaction effects.
For example, in the pure-state representation, a fully quantum treatment may be achieved by the functional
\begin{equation}\label{eq:quantum-corrected-energy}
	E
	={\int}D\bigg(\langle\widehat{H}\rangle+\frac{\hbar^2}{8mD^2}\big(2\|\nabla\widehat{P}\|^2-|\nabla D|^2\big)\bigg)\de\br,
\end{equation}
where we recall $D=\operatorname{Tr}\widehat{P}$ and the last two terms are associated with a combination of quantum potential and quantum geometric tensor\cite{PrVa80} contributions.
This form of potential energy has previously appeared in \citenum{FoHoTr19} in the context of nonadiabatic quantum hydrodynamics and evidently leads to the appearance of higher-order gradients.
The investigation of the fully quantum DDFT treatment arising from Eq.~\eqref{eq:quantum-corrected-energy} is left for future work and the remainder of this paper will restrict the analysis to the MQC Ehrenfest dynamics
\begin{equation}\label{eq:qces-hamiltonian-form}
	i\hbar\frac{\partial\widehat{P}}{\partial t}
	=i\hbar\mu\operatorname{div}\bigg(\widehat{P}\big\langle\nabla\widehat{H}\big\rangle+k_BT\frac{\widehat{P}}D\nabla D\bigg)+\big[\widehat{H},\widehat{P}\big]
\end{equation}
resulting from the total energy functional in Eq.~\eqref{eq:ehrenfest-energy}.

We conclude this section with a brief discussion of the possible limitations of the present Ehrenfest treatment.
For example, the Ehrenfest model completely fails for pure-dephasing dynamics since in that case the classical evolution decouples entirely.\cite{Franco,Manfredi23}
This drawback is shared by the proposed model.
In addition, standard nonadiabatic Ehrenfest dynamics is known to show significant limitations in the description of electronic coherences.
Similar issues may also arise in the overdamped regime considered here, although the present setting differs substantially from the full energy-conserving nonadiabatic dynamics problem.
In particular, the overall relaxation of the coupled system also damps electronic coherences and may therefore reduce the importance of features such as coherence revivals, which are typically more difficult to capture within Ehrenfest dynamics.
Whether this relaxation indeed mitigates the usual limitations of the Ehrenfest approximation remains an open question that will require a systematic comparison with alternative quantum and MQC approaches and is left for future work.

\subsection{Equilibrium states and detailed balance}
In this section we characterize the equilibrium states of QCES dynamics and show that its governing equation, Eq.~\eqref{eq:qces-hamiltonian-form}, retains detailed balance\cite{Mauri,Tully}.

We first construct a stationary state in the pure-state representation.
Let $\beta=(k_BT)^{-1}$ denote the inverse temperature.
In particular, we show that the Born--Oppenheimer equilibrium ansatz
\begin{equation}\label{eq:qces-pure-stationary}
	\widehat P_a
	=D_a|\phi_a\rangle\langle\phi_a|,
	\qquad\text{with}\qquad
	\widehat{H}|\phi_a\rangle
	=E_a|\phi_a\rangle,\qquad
	D_a
	=\frac{e^{-\beta E_a}}{\int e^{-\beta E_a}\,\de\br},
\end{equation}
yields steady states of Eq.~\eqref{eq:qces-hamiltonian-form}.
This follows from a direct application of the Hellmann--Feynman theorem.
Indeed, upon denoting $\langle\widehat{A}\rangle=\langle\phi_a|\widehat A|\phi_a\rangle$, we have $\langle\nabla\widehat H\rangle=\nabla E_a$.
Also, differentiating the Boltzmann density $D_a$ in Eq.~\eqref{eq:qces-pure-stationary} gives $\nabla D_a=-\beta D_a\nabla E_a$, so that the transport velocity in the equation for the density $D$ vanishes pointwise, that is $\boldsymbol v=-\mu\bigl[\langle\nabla\widehat H\rangle+\beta^{-1}\nabla\ln D_a\bigr]=0$.
In particular, the matrix flux $\boldsymbol v\widehat P_a=\boldsymbol v D_a|\phi_a\rangle\langle\phi_a|$ also vanishes.
Finally, $[\widehat H,\widehat P_a]=0$ and Eq.~\eqref{eq:qces-hamiltonian-form} therefore gives $\partial_t\widehat P_a=0$.

We now allow the conditional electronic state to be mixed and write $\widehat{P}=D\hat\rho$, where $\hat\rho(\br,t)$ is a conditional density matrix satisfying $\operatorname{Tr}\hat\rho(\br,t)=1$ at each point $\br$ in the reaction coordinate space.
It is easy to show that Eq.~\eqref{eq:qces-hamiltonian-form} drives the MQC state $\widehat{P}(\br,t)$ towards the equilibrium 
\begin{equation}\label{eq:conditional-gibbs-state}
	\widehat{P}_{\rm eq}(\br)
	=D_{\rm eq}(\br)\hat\rho_{\rm eq}(\br),
	\qquad\text{with}\qquad
	\hat\rho_{\rm eq}
	=\frac{e^{-\beta\widehat{H}}}{\operatorname{Tr}e^{-\beta\widehat{H}}},\qquad
	D_{\rm eq}
	=\frac{e^{-\langle\beta\widehat{H}+\ln\hat\rho_{\rm eq}\rangle}}{{\int}e^{-\langle\beta\widehat{H}+\ln\hat\rho_{\rm eq}\rangle}\de\br},
\end{equation}
and, in this case, $\langle\widehat{A}\rangle=\operatorname{Tr}(\hat\rho_{\rm eq}\widehat{A})$. 
Indeed, by the finite-temperature extension of the Hellmann--Feynman theorem\cite{PoEtAl20,Popov}, $\beta\langle\nabla\widehat{H}\rangle+\nabla\ln D_{\rm eq}=\nabla\big(\langle\beta\widehat{H}+\ln\hat\rho_{\rm eq}\rangle+\ln D_{\rm eq}\big)=0$ which, using Eq.~\eqref{eq:qces-hamiltonian-form}, leads directly to the conclusion that $\partial_t\widehat{P}_{\rm eq}=0$.
Importantly, since $\exp\langle-\beta\widehat{H}-\ln\hat\rho_{\rm eq}\rangle=\operatorname{Tr}e^{-\beta\widehat{H}}$, we have $D_{\rm eq}=\operatorname{Tr}e^{-\beta\widehat{H}}/\operatorname{Tr}{\int}e^{-\beta\widehat{H}}\de\br$ and therefore
\begin{equation}\label{eq:hybrid-gibbs-state}
	\widehat{P}_{\rm eq}
	=D_{\rm eq}\hat\rho_{\rm eq}
	=\frac{e^{-\beta\widehat{H}}}{\operatorname{Tr}{\int}e^{-\beta\widehat{H}}\de\br},
\end{equation}
thereby restoring detailed balance and addressing a long-standing debate in MQC dynamics\cite{Mauri,Tully}. 

These results are compatible with the corresponding results recently obtained in \citenum{TrMCGB} for a phase-space MQC Hamiltonian depending only on a set of Poisson-commuting coordinates (that is, $\br$ in the present case).
We emphasize that equilibrium profiles of the type above are instead unavailable for general phase-space MQC Hamiltonians, in which case detailed balance remains an outstanding question which lies outside the scope of the present work.

\section{Numerical implementation}\label{sec:num-implementation}
This section illustrates the computational schemes that were used in the numerical implementation of the QCES and QCLS equations.
In particular, we use atomic units and restrict to consider one spatial dimension so that $\hbar=1$ and the energy functional in Eq.~\eqref{eq:ehrenfest-energy} gives the QCES equation in Eq.~\eqref{eq:qces} the form
\begin{equation}\label{eq:qces-one-dimensional}
	\frac{\partial\widehat{P}}{\partial t}
	=\mu\frac{\partial}{\partial r}\bigg(\widehat{P}\bigg\langle\frac{\partial\widehat{H}}{\partial r}\bigg\rangle+k_BT\frac{\widehat{P}}D\frac{\partial D}{\partial r}\bigg)-i\big[\widehat{H},\widehat{P}\big],	
\end{equation}
while Eq.~\eqref{eq:qcls-free-energy-form} for QCLS reads
\begin{equation}\label{eq:qcls-one-dimensional}
	\frac{\partial\widehat{P}}{\partial t}
	=\mu\frac{\partial}{\partial r}\bigg(\widehat{P}\bigg\langle\frac{\partial\widehat{H}}{\partial r}\bigg\rangle+k_BT\frac{\partial\widehat{P}}{\partial r}\bigg)-i\big[\widehat{H},\widehat{P}\big].	
\end{equation}
We use exact local quantum propagation and conservative finite-volume transport to discretize both models in the case of two-level systems, so that the Hamiltonian $\widehat{H}(r)$ is a $2\times 2$ Hermitian matrix.
For the QCES model, the evolution is written in terms of the total transport velocity
\begin{equation}\label{eq:qces-transport-velocity}
	v
	=-\mu a-\kappa\frac{\partial_rD}{D}
\end{equation}
as $\partial_t\widehat P=-i[\widehat H,\widehat P]-\partial_r(v\widehat P)$, where $\kappa=\mu k_BT$ and $a=\operatorname{Tr}(\widehat P\partial_r\widehat H)/D$.

\subsection{Spatial grid and quantum propagation}
On a finite box $[R_{\min},R_{\max}]$, we introduce the nodes
\begin{equation}\label{eq:spatial-grid}
	r_j
	=R_{\min}+j\mathsf{h},\qquad
	\mathsf{h}
	=\frac{R_{\max}-R_{\min}}{N},\qquad
	j
	=0,\ldots,N,	
\end{equation}
and define the grid operators $\widehat P_j\approx\widehat P(r_j)$, $\widehat H_j=\widehat H(r_j)$, and $\widehat H'_j=\partial_r\widehat H(r_j)$.
We write $\widehat P_{\mathsf{h}}=(\widehat P_j)_{j=0}^N$ for the collection of grid operators.

Each grid node is assigned a cell.
Cell boundaries lie halfway between neighboring nodes, with the two outer boundaries at $R_{\min}$ and $R_{\max}$.
The cell widths are therefore
\begin{equation}\label{eq:cell-widths}
	w_0
	=w_N
	=\frac{\mathsf{h}}{2},\qquad
	w_j
	=\mathsf{h},\quad
	1\leq j\leq N-1.
\end{equation}
Thus, $M_{\mathsf{h}}(\widehat P_{\mathsf{h}}):=\sum_{j=0}^Nw_j\operatorname{Tr}\widehat P_j$ is the trapezoidal approximation of the total probability mass of the classical density $D$.
We impose no-flux boundary conditions, which prevent probability mass from entering or leaving the computational interval.
All calculations presented in this paper use $R_{\min}=-10$, $R_{\max}=30$, and $N=2^{12}=4096$ intervals, together with $k_BT=25.9\,\mathrm{meV}$ ($T=300\,\mathrm{K}$).
These numerical settings are identical for all solvents and for both the ground- and excited-state dynamics discussed in Section~\ref{sec:results}.
The interval $[R_{\min},R_{\max}]$ is chosen such that the boundary densities remain negligible throughout: over all simulations, $D$ never exceeds $1.2\cdot10^{-6}$ at $R_{\min}$, a value attained by the initial thermal distribution, which lies about five standard deviations inside the interval, and remains at the level of round-off ($\sim 10^{-14}$) at $R_{\max}$.

Standard Strang splitting\cite{Strang1968} separates an evolution equation into two subproblems and composes their flows in a symmetric sequence.
Here, we use this composition to separate the QC dynamics into exact local quantum evolution and classical transport and call the resulting step of length $s>0$ the \emph{quantum-classical Strang splitting}:
\begin{equation}\label{eq:quantum-classical-splitting}
	\mathcal S_s
	=\mathcal Q_{s/2}\circ\mathcal C_s\circ\mathcal Q_{s/2},\qquad
	(\mathcal Q_s\widehat P_{\mathsf{h}})_j
	=\widehat U_j(s)\widehat P_j\widehat U_j(s)^\dagger.
\end{equation}
Here, $\mathcal Q_s$ denotes the exact local quantum flow generated by $\widehat U_j(s)=\exp(-is\widehat H_j)$, with $\dagger$ denoting the Hermitian adjoint, while $\mathcal C_s$ denotes the discrete classical transport step.
In particular, for the two-level Hamiltonian, we write $\widehat H_j=h_0(r_j)\widehat{\sigma}_0+\boldsymbol h(r_j)\cdot\widehat{\boldsymbol\sigma}$, with $\widehat{\boldsymbol\sigma}$ the vector of Pauli matrices and $\widehat{\sigma}_0$ the identity matrix.
The Pauli coefficients $h_0,\ldots,h_3$ are real-valued functions of $r$, with $\boldsymbol h(r)=(h_1(r),h_2(r),h_3(r))$.
Then, the local evolution operators are $\widehat U_j(s)=\big[\cos(s|\boldsymbol h(r_j)|)\widehat{\sigma}_0-i|\boldsymbol h(r_j)|^{-1}{\sin(s|\boldsymbol h(r_j)|)}{\boldsymbol h(r_j)\cdot\widehat{\boldsymbol\sigma}}\big]\exp({-is h_0(r_j)})$, with the sine quotient interpreted by continuity at $|\boldsymbol h(r_j)|=0$.
The scalar phase cancels in the conjugation and can be omitted computationally.
This update preserves the local trace, Hermiticity and positive semidefiniteness.

\subsection{Implicit Ehrenfest--Smoluchowski transport}
For a classical substep, let $\widehat P_{\mathsf{h}}^{\mathrm{in}}$ denote the state after the first quantum half-step and $\widehat P_{\mathsf{h}}^{\mathrm{out}}=\mathcal C_s\widehat P_{\mathsf{h}}^{\mathrm{in}}$.
Classical coefficients are evaluated with the real duality pairing:
\begin{equation}\label{eq:real-duality-pairing}
	\langle\widehat A,\widehat B\rangle
	=\operatorname{Re}\operatorname{Tr}(\widehat A^\dagger\widehat B),\qquad
	D_j^{\mathrm{in}}
	=\langle\widehat{\sigma}_0,\widehat P_j^{\mathrm{in}}\rangle,\qquad
	a_j
	=\frac{\langle\widehat H'_j,\widehat P_j^{\mathrm{in}}\rangle}{D_j^{\mathrm{in}}}.
\end{equation}
For Hermitian $\widehat P_j^{\mathrm{in}}$ and $\widehat H'_j$, these expressions agree with the usual traces.
At the start of each classical substep, we evaluate $a_j$ from the input state, setting $a_j=0$ wherever $D_j^{\mathrm{in}}\leq10^{-14}$ to regularize small density tails, and freeze $a_j$ during the substep.

We discretize the scalar current $J=Dv$ using upwind differences for the advective contribution and centered differences for the diffusive contribution.\cite{Iserles2009}
At each interior face, define the force-driven drift and scalar transport coefficients:
\begin{equation}\label{eq:scalar-flux-coefficients}
	\begin{aligned}
		b_{j+1/2}
		&=-\frac{\mu}{2}(a_j+a_{j+1}),\\
		\alpha_{j+1/2}
		&=b_{j+1/2}^++\frac{\kappa}{\mathsf{h}}
		\geq 0,
		&\beta_{j+1/2}&=b_{j+1/2}^--\frac{\kappa}{\mathsf{h}}
		\leq 0,
	\end{aligned}
\end{equation}
where $z^+=\max(z,0)$ and $z^-=\min(z,0)$.
The output scalar density is obtained by backward Euler:
\begin{equation}\label{eq:scalar-density-update}
	\begin{aligned}
		\frac{D_j^{\mathrm{out}}-D_j^{\mathrm{in}}}{s}
		&=-\frac{J_{j+1/2}-J_{j-1/2}}{w_j},\\
		J_{j+1/2}
		&=\alpha_{j+1/2}D_j^{\mathrm{out}}+\beta_{j+1/2}D_{j+1}^{\mathrm{out}},
	\end{aligned}
\end{equation}
with boundary conditions $J_{-1/2}=J_{N+1/2}=0$.
This yields a tridiagonal linear system, which can be solved efficiently by the Thomas algorithm.\cite{Higham2002}

Once the output densities (and therefore also the scalar currents) are known, we define the total face velocity using the density in the donating cell:
\begin{equation}\label{eq:face-velocity}
	v_{j+1/2}
	=
	\begin{cases}
		J_{j+1/2}/D_j^{\mathrm{out}}, & J_{j+1/2}>0,\\
		J_{j+1/2}/D_{j+1}^{\mathrm{out}}, & J_{j+1/2}<0,\\
		0, & J_{j+1/2}=0.
	\end{cases}
\end{equation}
For non-negative densities, the total face velocity is well-defined even when a cell is empty.
Indeed, the signs of $\alpha_{j+1/2}$ and $\beta_{j+1/2}$ imply:
\begin{equation}\label{eq:empty-cell-flux}
	\begin{aligned}
		D_j^{\mathrm{out}}=0
		&\;\Rightarrow\;
		J_{j+1/2}
		=\beta_{j+1/2}D_{j+1}^{\mathrm{out}}\leq 0,\\
		D_{j+1}^{\mathrm{out}}
		=0
		&\;\Rightarrow\;
		J_{j+1/2}
		=\alpha_{j+1/2}D_j^{\mathrm{out}}\geq 0.
	\end{aligned}
\end{equation}
Thus, each branch with nonzero current has a strictly positive denominator, namely the density in the donating cell.

We discretize the matrix-valued transport equation $\partial_t\widehat P=-\partial_r(v\widehat P)$ using the same conservative finite-volume method and backward Euler.
Here, $\widehat J_{j+1/2}$ denotes the matrix-valued numerical flux approximating $v\widehat P$ at the cell interface $r_{j+1/2}$, evaluated by
\begin{equation}\label{eq:matrix-density-update}
	\begin{aligned}
		\frac{\widehat P_j^{\mathrm{out}}-\widehat P_j^{\mathrm{in}}}{s}
		&=-\frac{\widehat J_{j+1/2}-\widehat J_{j-1/2}}{w_j},
		\qquad\qquad
		\widehat J_{j+1/2}
		=v_{j+1/2}^+\widehat P_j^{\mathrm{out}}+v_{j+1/2}^-\widehat P_{j+1}^{\mathrm{out}},
	\end{aligned}
\end{equation}
with $\widehat J_{-1/2}=\widehat J_{N+1/2}=0$.
A positive current carries the output quantum state of the left cell; a negative current carries that of the right cell.
All four complex matrix entries are evolved using the same tridiagonal coefficient matrix.
After multiplication by $W=\operatorname{diag}(w_0,\ldots,w_N)$, the coefficient matrices of the scalar density system in Eq.~\eqref{eq:scalar-density-update} and the operator-valued transport system in Eq.~\eqref{eq:matrix-density-update} have positive diagonal entries and non-positive off-diagonal entries.
Moreover, conservation of the numerical fluxes implies that their column sums equal the positive quadrature weights $w_j$.
Thus, both matrices are strictly column diagonally dominant and therefore nonsingular $M$-matrices.\cite{BermanPlemmons1994}
In particular, their inverses are entrywise non-negative.
Consequently, each output operator is a non-negative linear combination of the input operators, preserving Hermiticity and positive semidefiniteness.

We note that the scalar and matrix densities remain consistent at every node.
Indeed, the face velocity satisfies $v_{j+1/2}^{+}D_j^{\mathrm{out}}+v_{j+1/2}^{-}D_{j+1}^{\mathrm{out}}=J_{j+1/2}$.
Thus, the output densities satisfy the equation obtained by taking the trace of the matrix update, and uniqueness gives $\operatorname{Tr}\widehat P_j^{\mathrm{out}}=D_j^{\mathrm{out}}$.
Moreover, the conservative form of the matrix update ensures conservation of the total probability mass.
Multiplying the matrix update by $w_j$, summing over $j$, and using that the boundary fluxes vanish yields $M_{\mathsf{h}}(\widehat P_{\mathsf{h}}^{\mathrm{out}})=M_{\mathsf{h}}(\widehat P_{\mathsf{h}}^{\mathrm{in}})$.
Since the quantum substeps preserve the local trace, this mass conservation also holds for the full QC Strang splitting, up to round-off.

\subsection{Liouville--Smoluchowski transport}
For the Liouville--Smoluchowski model, diffusion acts directly on the matrix density, giving $\widehat J^{\scriptscriptstyle\mathrm{QCLS}}=-\mu\widehat P a-\kappa\partial_r\widehat P$.
With the same frozen force and backward Euler update, the discrete flux is:
\begin{equation}\label{eq:qcls-matrix-flux}
	\widehat J_{j+1/2}^{\scriptscriptstyle\mathrm{QCLS}}
	=\alpha_{j+1/2}\widehat P_j^{\mathrm{out}}+\beta_{j+1/2}\widehat P_{j+1}^{\mathrm{out}}.
\end{equation}
Substituting this into the balance equation in Eq.~\eqref{eq:matrix-density-update} gives one tridiagonal solve for all matrix entries.
The same argument establishes preservation of Hermiticity, positive semidefiniteness, and conservation of the total probability mass.

\subsection{Time-step control}
For a fixed time step $s$, the update would be $\widehat P_{\mathsf{h}}^{n+1}=\mathcal S_s(\widehat P_{\mathsf{h}}^n)$.
We use adaptive time stepping based on step doubling for ground-state relaxation and for the excited-state dynamics that provides the initial states at the emission delays.
The absolute local tolerance is $\varepsilon_{\mathrm{loc}}=10^{-8}$.
For each trial step of length $s$, we compare one full step with two half-steps:
\begin{equation}\label{eq:step-doubling-error}
	\widehat P_{\mathsf{h}}^{\mathrm{coarse}}
	=\mathcal S_s(\widehat P_{\mathsf{h}}^n),\qquad
	\widehat P_{\mathsf{h}}^{\mathrm{fine}}
	=\mathcal S_{s/2}\left(\mathcal S_{s/2}(\widehat P_{\mathsf{h}}^n)\right),\qquad
	e
	=\max_{j,\ell,m}\left|(\widehat P_j^{\mathrm{fine}}-\widehat P_j^{\mathrm{coarse}})_{\ell m}\right|.
\end{equation}
When $e\leq\varepsilon_{\mathrm{loc}}$, we accept $\widehat P_{\mathsf{h}}^{n+1}=\widehat P_{\mathsf{h}}^{\mathrm{fine}}$ and advance time by $s$.
Otherwise, the trial is repeated from $\widehat P_{\mathsf{h}}^n$ with a smaller step.
The next trial step size is chosen as $\lambda s$, with the multiplicative factor $\lambda=0.9(\varepsilon_{\mathrm{loc}}/e)^{1/2}$ restricted to $[0.2,5]$ and the step capped at $s_{\max}=50\,\mathrm{a.u.}$.
For $e=0$, we use the maximum growth factor $\lambda=5$.
For the short-time insets in Fig.~\ref{fig:equilibrium} (first $30\,\mathrm{fs}$), we use $\varepsilon_{\mathrm{loc}}=10^{-10}$ and $s_{\max}=0.1\,\mathrm{a.u.}$ to resolve the electronic oscillations, and the spectral generating functions of Section~\ref{sec:spectra} are propagated with fixed steps $s=0.1\,\mathrm{a.u.}$.
We note that the safety factor $0.9$ reduces the predicted step, leaving a margin for errors in the prediction, and the bounds $0.2$ and $5$ limit each controller adjustment to a fivefold decrease or increase, preventing abrupt changes in step size.
These safeguards follow standard bounded step-size control.\cite{HairerNorsettWanner1993}

\section{Solvation model and simulation results}\label{sec:results}
\subsection{The solute-solvent coupled model}
\begin{figure}[t!]
	\centering
	\includegraphics[width=\textwidth]{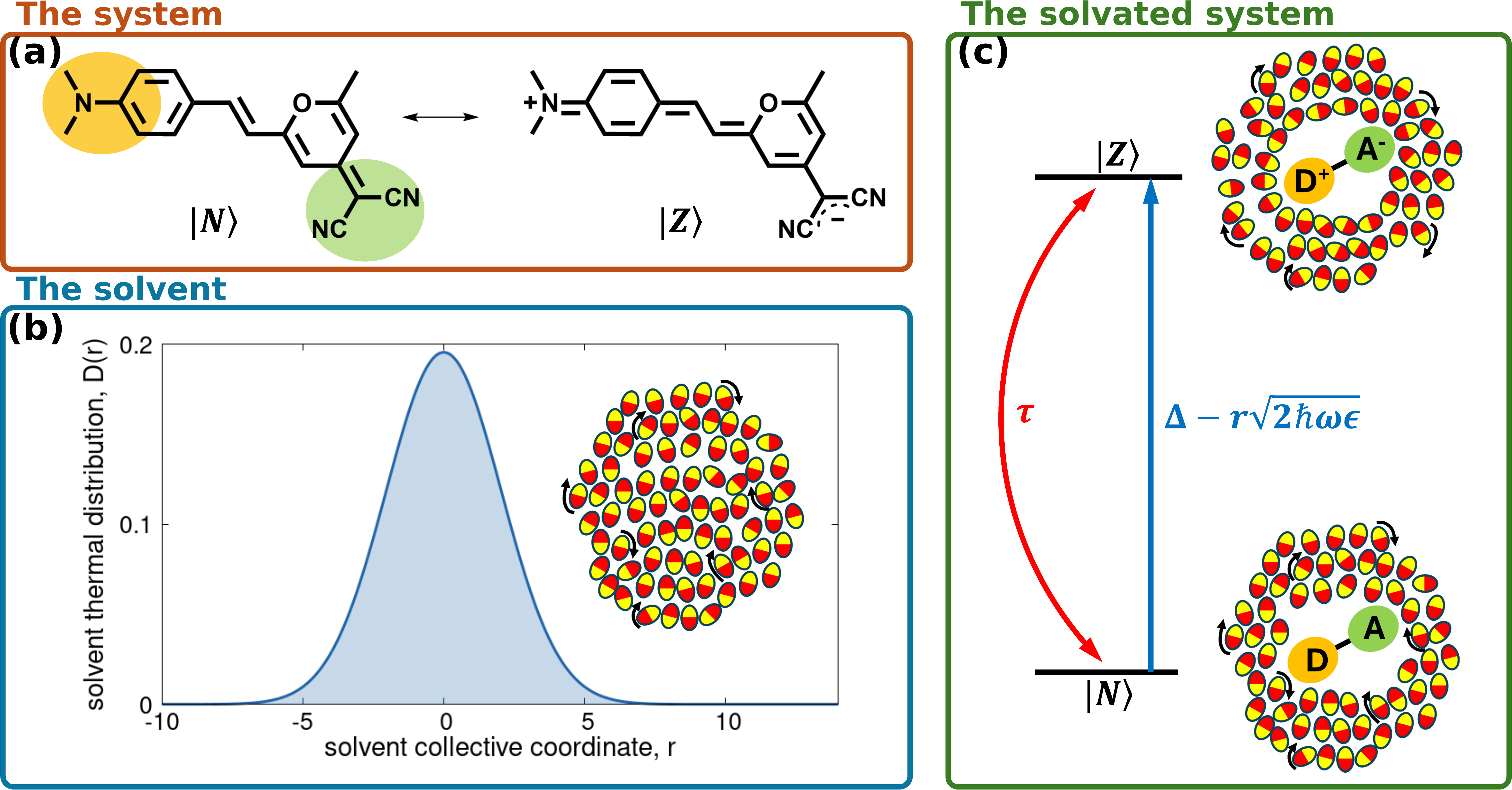}
	\caption{Model of the DCM dye in a dipolar solvent.
	(a) Chemical structure of DCM \mbox{(4-(dicyanomethylene)-2-methyl-6-(4-dimethylaminostyryl)-4H-pyran)} with its neutral, $|N\rangle$, and zwitterionic, $|Z\rangle$, resonance structures.
	Donor (orange) and acceptor (green) moieties highlighted.
	(b) Thermal distribution, $D(r)$, of the classical solvent coordinate~$r$.
	Inset: solvent shell of reorienting dipolar molecules (red/yellow poles).
	(c) Two-level scheme for the solvated DCM dye; side cartoons show the solvent shell around the neutral and zwitterionic states.
	Notice that the energy gap between $|N\rangle$ and $|Z\rangle$ (blue arrow) is tuned by the coupling to the solvent coordinate $r$.}
	\label{fig:model}
\end{figure}

The low-energy electronic structure of polar conjugated dyes can be described within a minimal essential state model based on the resonance between a neutral state, $|N\rangle$, with negligible polarity, and a charge-separated (zwitterionic) state, $|Z\rangle$, carrying a sizable dipole moment $\mu_0$.\cite{Painelli1998}
Here, we consider the solvation process in three different solvents: chloroform ($\mathrm{CHCl_3}$), dichloromethane ($\mathrm{CH_2Cl_2}$), and dimethyl sulfoxide (DMSO).
This picture is illustrated in Fig.~\ref{fig:model}(a) for the representative dye DCM, a merocyanine-type donor-acceptor fluorophore.
The associated MQC Hamiltonian reads
\begin{equation}\label{eq:total-hamiltonian}
	\widehat H(r)
	=\frac{\hbar\omega}{2}r^2\widehat{\sigma}_0+\widehat H_{\mathrm{dye-solvent}}(r)+\widehat H_{\mathrm{dye}},
\end{equation}
where we have modeled the solvent as an elastic harmonic bath in the first term, while the second and third terms are the dye-solvent interaction and the uncoupled dye Hamiltonian, respectively.
The two diabatic states are separated by an energy gap $\Delta$ and coupled by the electronic mixing parameter $\tau$, so that the dye Hamiltonian reads:
\begin{equation}\label{eq:dye-hamiltonian}
	\widehat H_{\mathrm{dye}}
	=\tau\Big(|N\rangle\langle Z|+|Z\rangle\langle N|\Big)+\Delta\,|Z\rangle\langle Z|.	
\end{equation}
Despite its minimal form, this Hamiltonian captures the essential polarity and polarizability of the dye and has been successfully used as a basis for describing solvent effects on the optical properties of polar chromophores.\cite{Painelli1998,Comoretto2000,Sanyal2017,DiMaiolo2018,Giavazzi2022,Giavazzi2023}

The solvation process is governed by the coupling between the dye and the solvent reaction field.\cite{Painelli1999}
The fast electronic contribution of the solvent is treated in the antiadiabatic approximation and is therefore absorbed into a renormalization of the molecular parameters.\cite{Phan2020}
Since the refractive index varies only moderately among common organic solvents, these renormalized parameters are taken to be solvent independent.
The slower orientational contribution is described explicitly through the one-dimensional, dimensionless reaction-field coordinate $r$ introduced in Section~\ref{sec:intro}.
Its coupling to the dye is introduced through:
\begin{equation}\label{eq:dye-solvent-interaction}
	\widehat H_{\mathrm{dye-solvent}}(r)
	=-r\sqrt{2\hbar\omega\epsilon}\,|Z\rangle\langle Z|,
\end{equation}
where $\epsilon$ is the solvent relaxation energy, which measures the strength of the solute-solvent coupling, and $\hbar\omega$ identifies the characteristic energy scale of the solvent coordinate.
Therefore, in the ordered diabatic basis $\{|N\rangle,|Z\rangle\}$, the solute-solvent Hamiltonian in Eq.~\eqref{eq:total-hamiltonian} reads
\begin{equation}\label{eq:solute-solvent-hamiltonian}
	\widehat H(r)
	=
	\begin{pmatrix}
		\frac{\hbar\omega}{2}r^2 & \tau\\
		\tau & \frac{\hbar\omega}{2}r^2-\alpha r+\Delta
	\end{pmatrix}
	=
	h_0(r)\widehat{\sigma}_0+{h}_1(r)\widehat{\sigma}_1+{h}_3(r)\widehat{\sigma}_3,
\end{equation}
where $\alpha=\sqrt{2\hbar\omega\epsilon}$, $\widehat{\sigma}_1=|N\rangle\langle Z|+|Z\rangle\langle N|$, $\widehat{\sigma}_3=|N\rangle\langle N|-|Z\rangle\langle Z|$, and
\begin{equation}\label{eq:pauli-coefficients}
	h_0(r)
	=\frac12\big({\hbar\omega}r^2-{\alpha}r+{\Delta}\big),\qquad\quad
	h_1(r)
	=\tau,\qquad\quad
	h_3(r)
	=\frac12\big({\alpha}r-{\Delta}).
\end{equation}
The Hamiltonian in Eq.~\eqref{eq:solute-solvent-hamiltonian} can be diagonalized exactly at fixed values of the reaction field, yielding the adiabatic potential energy curves of the coupled solute-solvent problem.

In our approach, the classical solvent distribution $D(r,t)$ (Fig.~\ref{fig:model}(b)) evolves dynamically within the MQC framework developed above, while the quantum state of the dye is given by $\hat\varrho(t)=\int\widehat P(r,t)\,\de r$.
The quantum-classical correlations are incorporated by the conditional density matrix $\hat\rho(r,t)=\widehat P(r,t)/D(r,t)$.

The dye parameters are fixed at $\Delta=2.28\,\mathrm{eV}$ and $\tau=0.88\,\mathrm{eV}$ for all solvents.\cite{Boldrini2002}
The dye-solvent relaxation energy $\epsilon$ is $0.32\,\mathrm{eV}$ in $\mathrm{CHCl_3}$, $0.35\,\mathrm{eV}$ in $\mathrm{CH_2Cl_2}$, and $0.75\,\mathrm{eV}$ in DMSO.
The corresponding solvent longitudinal relaxation times $\tau_L$ are $2.8$, $0.56$, and $2.0\,\mathrm{ps}$, respectively.\cite{Horng1995}
In the overdamped limit of the orientational solvent dynamics, these relaxation times are related to the orientational friction coefficient and characteristic frequency through $\tau_L=\zeta/\omega^2$.
They therefore characterize the slow, friction-dominated solvent relaxation described by the present Smoluchowski model.
These determine the solvent mobility as 
\begin{equation}\label{eq:solvent-mobility}
	\mu
	=\frac{1}{\hbar\omega\,\tau_L}.
\end{equation}
We use $\hbar\omega=6.20\,\mathrm{meV}$ for all the considered solvents, which corresponds to a wavenumber of $50\,\mathrm{cm}^{-1}$.
Since the solvent coordinate is treated in the overdamped regime, the precise value of $\omega$ is not expected to play a major dynamical role, provided that it remains sufficiently low to represent a thermally populated solvent mode.
The resulting mobilities are $57.6$, $288.7$, and $80.6\,\mathrm{eV}^{-1}\,\mathrm{ps}^{-1}$ for $\mathrm{CHCl_3}$, $\mathrm{CH_2Cl_2}$, and DMSO, respectively.
Using nearby values of $\omega$, while adjusting $\mu$ so as to preserve the corresponding $\tau_L$, does not appreciably affect the results.

The dynamics is initialized from an uncorrelated solute-solvent state in which the dye is localized in the neutral diabatic state $|N\rangle$, while the solvent coordinate follows the thermal Gaussian distribution associated with the harmonic potential (Fig.~\ref{fig:model}(b)).
Accordingly, the initial condition reads:
\begin{equation}\label{eq:neutral-initial-state}
	\widehat P_0(r)
	=D_0(r)|N\rangle\langle N|,
	\qquad\text{with}\qquad
	D_0(r)
	=\sqrt{\frac{\hbar\omega}{2\pi k_BT}}\,\exp\left(-\frac{\hbar\omega}{2k_BT}r^2\right).
\end{equation}
Although the initial condition is factorized, the subsequent coupled dynamics generally builds correlations between the electronic state and the solvent coordinate, so that $\widehat P(r,t)$ is no longer factorized as $D(r,t)\hat\varrho(t)$ at later times.

\subsection{Ground-state thermalization}
\begin{figure*}[!t]
	\centering
	\includegraphics[width=\textwidth]{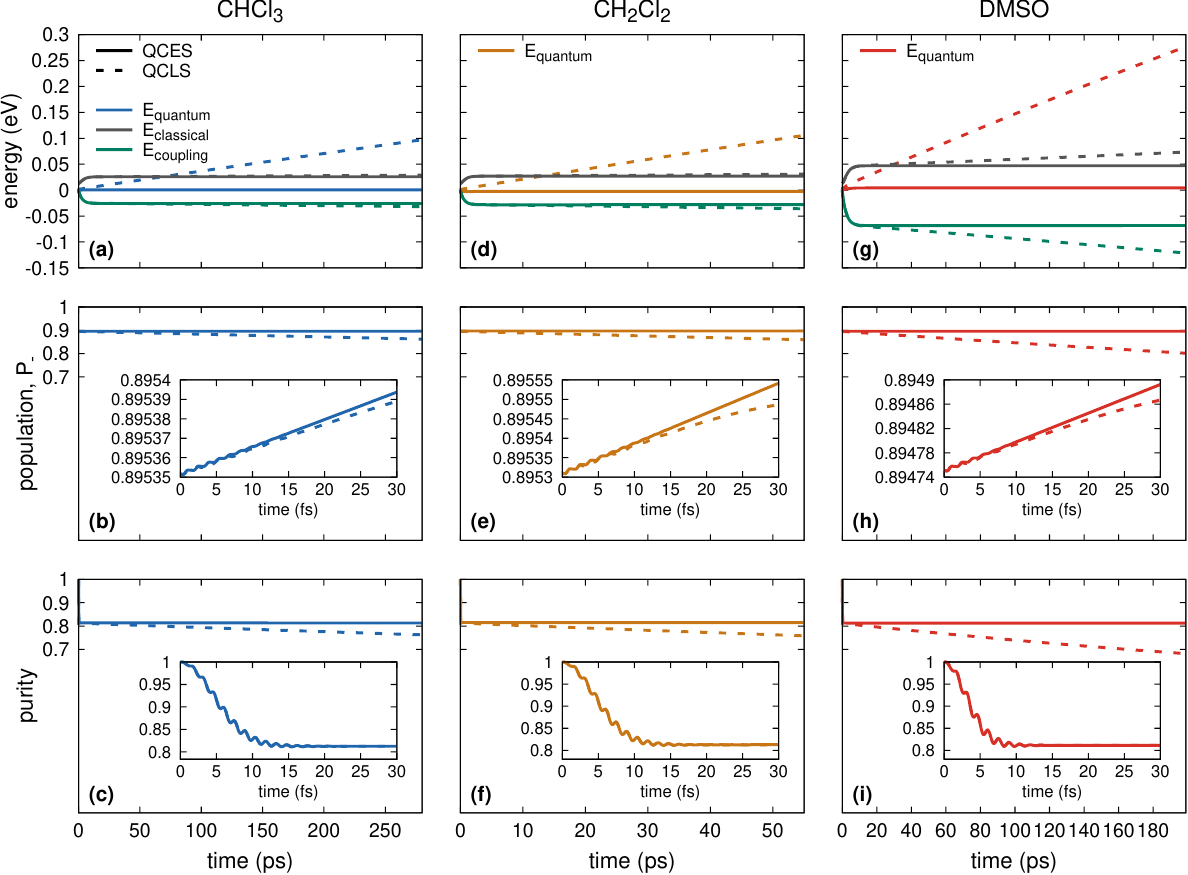}
	\caption{Comparison of the ground state dynamics obtained with the Ehrenfest--Smoluchowski equation (solid lines) and the Liouville--Smoluchowski equation (dashed lines) for CHCl$_3$ (panels (a), (b), and (c)), CH$_2$Cl$_2$ (panels (d), (e), and (f)), and DMSO (panels (g), (h), and (i)).
	Panels (a), (d), and (g) show the quantum energy $E_{\text{quantum}}$, classical solvent energy $E_{\text{classical}}$, and solute-solvent coupling energy $E_{\text{coupling}}$.
	Panels (b), (e), and (h) show the population $P_-$ of the lower adiabatic state, while panels (c), (f), and (i) show the purity; insets highlight the first $30\,\mathrm{fs}$ of the dynamics.
	In the purity insets, the QCES and QCLS curves are indistinguishable on the given scale.
	Quantum energies are shown in blue, orange, and red for CHCl$_3$, CH$_2$Cl$_2$, and DMSO, respectively, while the classical and coupling energies are shown in grey and green.
	Model parameters are reported in the main text.}
	\label{fig:equilibrium}
\end{figure*}

We first examine the relaxation toward thermal equilibrium predicted by the two MQC Smoluchowski equations, Eq.~\eqref{eq:qcls} and Eq.~\eqref{eq:qces}.
Figure~\ref{fig:equilibrium} compares the dynamics of DCM in CHCl$_3$, CH$_2$Cl$_2$, and DMSO, starting from the factorized initial condition introduced above.
Solid curves show results obtained with the QCES equation, whereas dashed curves show those obtained with the QCLS equation.
The three energy components are
\begin{align}
	E_{\rm classical}(t)
	&=\frac{\hbar\omega}{2}{\int} r^2D(r,t)\,\de r,\label{eq:classical-solvent-energy}\\
	E_{\rm coupling}(t)
	&=-\alpha{\int} r\,\langle Z|\widehat P(r,t)|Z\rangle\,\de r,\label{eq:solute-solvent-coupling-energy}\\
	E_{\rm quantum}(t)
	&=\operatorname{Tr}{\int}\widehat H_{\rm dye}\widehat P(r,t)\,\de r,\label{eq:quantum-dye-energy}
\end{align}
whose sum is the total energy.
The population in the plots is that of the lower adiabatic state, that is $P_-(t)$ with the notation
\begin{equation}\label{eq:adiabatic-populations}
	P_\pm(t)
	=\int\langle\phi_\pm(r)|\widehat P(r,t)|\phi_\pm(r)\rangle\,\de r
	=1-P_\mp(t),
\end{equation}
where $|\phi_\pm(r)\rangle$ is the upper/lower eigenstate of the solute-solvent Hamiltonian at fixed $r$. 
Since the initial diabatic state $|N\rangle$ is not an eigenstate of the solute-solvent Hamiltonian, $P_-$ starts at $0.895$ rather than at one.
The electronic purity is defined as $\mathcal{P}(t)=\operatorname{Tr}[\hat{\varrho}(t)^2]$, so that $\mathcal{P}=1$ corresponds to a pure electronic state whereas smaller values indicate that the electronic state becomes mixed and thus correlated with the solvent coordinate.

For DCM in CHCl$_3$ ($\epsilon=0.32\,\mathrm{eV}$, $\tau_L=2.8\,\mathrm{ps}$), Fig.~\ref{fig:equilibrium}(a) reports the quantum energy of the dye, the classical solvent energy, and the solute-solvent coupling energy over the first $280\,\mathrm{ps}$ of the dynamics.
Within the QCES description, all three contributions rapidly approach stationary values.
After approximately $20\,\mathrm{ps}$, the classical solvent energy reaches about $26\,\mathrm{meV}$, reflecting the reorganization of the solvent around the solute.
This energetic cost is compensated by a stabilizing solute-solvent coupling energy of approximately $-26\,\mathrm{meV}$.
The quantum energy of the dye settles within the first few picoseconds at about $1\,\mathrm{meV}$.
Consistently, the electronic state remains predominantly in the lower adiabatic state, with a population of about $90\%$ at equilibrium, as shown in Fig.~\ref{fig:equilibrium}(b).
The electronic purity, shown in Fig.~\ref{fig:equilibrium}(c), decreases from its initial value $\mathcal P=1$ to about $0.81$ within the first $15\,\mathrm{fs}$, displaying damped oscillations at a frequency set by the electronic energy gap.
It then remains nearly constant for the rest of the simulation.
This loss of purity reflects the build-up of correlations between the electronic state and the solvent coordinate, after which the coupled solute-solvent system has reached a stationary state.

A markedly different behavior is obtained with the QCLS model.
The dashed curves in panel (a) do not approach stationary values and instead develop a divergent behavior at long times.
This instability is accompanied by a progressive decrease of the lower adiabatic population (panel (b)), indicating a continuous transfer of population to the upper adiabatic state.
As the electronic state departs from the stationary solution obtained with the QCES equation, the classical solvent and solute-solvent coupling energies also progressively deviate from their equilibrium values.
The electronic purity follows the same trend (panel (c)): during the first $30\,\mathrm{fs}$, the QCLS and QCES results are essentially indistinguishable, showing the same initial decay and short-time oscillatory behavior.
At longer times, however, the QCLS purity does not remain stationary but drifts slowly to lower values, reaching about $0.76$ after $280\,\mathrm{ps}$.

The same qualitative picture is obtained for DCM in CH$_2$Cl$_2$ ($\epsilon=0.35\,\mathrm{eV}$, $\tau_L=0.56\,\mathrm{ps}$), as shown in panels (d), (e), and (f).
Because the solvent relaxation energy is close to that of CHCl$_3$, the equilibrium energy partitioning is similar.
The much shorter longitudinal relaxation time, however, leads to a correspondingly faster approach to the stationary state: within the QCES dynamics, the quantum, classical, and coupling energies have essentially stabilized after about $4\,\mathrm{ps}$.
The QCLS model again fails to reach a stationary thermal state and develops the same long-time instability.

The effect becomes even more pronounced for DCM in DMSO ($\epsilon=0.75\,\mathrm{eV}$, $\tau_L=2\,\mathrm{ps}$), for which the solute-solvent coupling is substantially stronger (Fig.~\ref{fig:equilibrium}(g), (h), and (i)).
Within the QCES description, the classical solvent energy approaches approximately $47\,\mathrm{meV}$ after about $15\,\mathrm{ps}$, while the solute-solvent coupling energy reaches about $-68\,\mathrm{meV}$.
The quantum contribution remains close to zero, consistent with an electronic state that is still predominantly described by the neutral $|N\rangle$ configuration.
The electronic purity likewise reaches a stationary value of about $0.81$, which is maintained up to $199\,\mathrm{ps}$.
In contrast, the QCLS equation again fails to produce thermalization: the energy components progressively depart from stationary values as the solvent-induced admixture of the $|Z\rangle$ state continues to increase.
Correspondingly, the electronic purity keeps decreasing, reaching about $0.68$ after $199\,\mathrm{ps}$, well below the stationary value obtained with the QCES equation.

These results show that the newly derived QCES equation drives the coupled dye-solvent dynamics toward a stationary thermal state for all three solvents considered, whereas the QCLS equation fails to retain stable long-time relaxation.

\subsection{Fine-structure solvent dynamics}\label{sec:shape}
\begin{figure*}[t!]
	\centering
	\includegraphics[width=\textwidth]{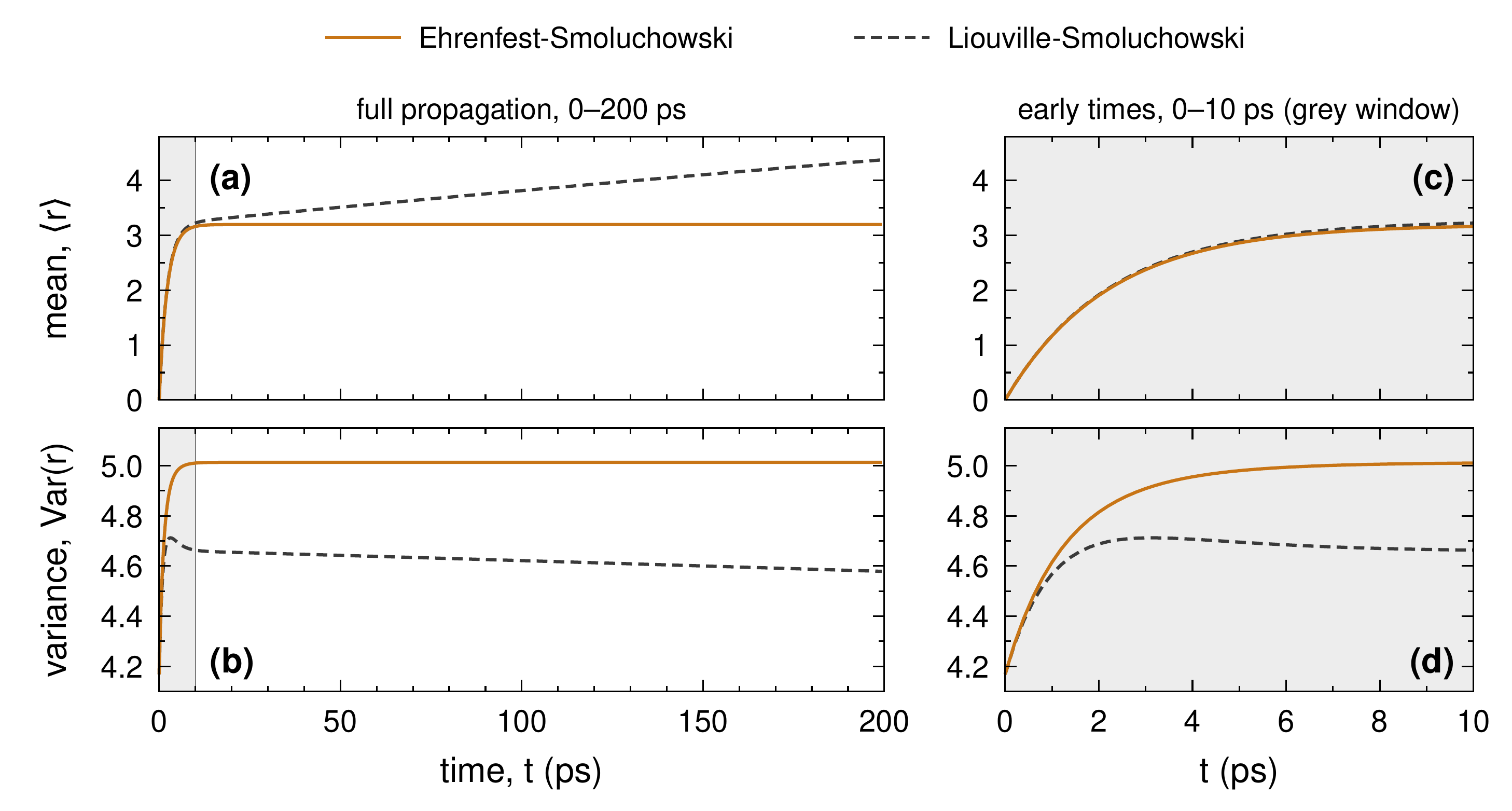}
	\caption{Mean $\langle r\rangle(t)$ (panels (a) and (c)) and variance $\mathrm{Var}(r)(t)$ (panels (b) and (d)) of the solvent distribution for DCM in DMSO, obtained with the Ehrenfest--Smoluchowski equation (solid lines) and the Liouville--Smoluchowski equation (dashed lines).
	Panels (a) and (b) show the full propagation up to $200\,\mathrm{ps}$; panels (c) and (d) magnify the first $10\,\mathrm{ps}$ (grey window).
	The initial condition is the factorized state with the dye in $|N\rangle$.}
	\label{fig:moments}
\end{figure*}
The energies, populations, and purity discussed above are integrated quantities.
This section presents a more detailed study of the two MQC Smoluchowski equations in terms of the solvent distribution $D(r,t)$.
Figure~\ref{fig:moments} shows its mean and variance,
\begin{equation}\label{eq:solvent-mean-variance}
	\langle r\rangle(t)
	=\int r\,D(r,t)\,\de r,\qquad
	\mathrm{Var}(r)(t)
	=\int\big(r-\langle r\rangle(t)\big)^2D(r,t)\,\de r,
\end{equation}
for DCM in DMSO, the solvent with the strongest solute-solvent coupling.
During the first 10 picoseconds, the two descriptions show an almost identical shift of the distribution towards larger positive values of $r$ (panel (c)).
Its width, however, evolves differently already after about $1\,\mathrm{ps}$ (panel (d)): the QCES variance increases monotonically, whereas the QCLS variance passes through a maximum and decreases afterwards.
On longer timescales the difference becomes qualitative.
Both QCES moments reach stationary values after about $20\,\mathrm{ps}$, $\langle r\rangle\approx 3.2$ and $\mathrm{Var}(r)\approx 5.0$.
In the QCLS dynamics, instead, the mean keeps increasing almost linearly up to $\langle r\rangle\approx 4.4$ at $200\,\mathrm{ps}$ (panel (a)), while the variance continues to decrease (panel (b)) towards an increasingly peaked profile.
The resulting solvent drift is the origin of the long-time instability of the energy components discussed in the previous section.
The same scenario is found for CHCl$_3$ and CH$_2$Cl$_2$, where the QCES mean becomes stationary at $\langle r\rangle\approx 1.97$ and $2.05$, respectively, while the QCLS mean continues to grow.

\begin{figure*}[t!]
	\centering
	\includegraphics[width=\textwidth]{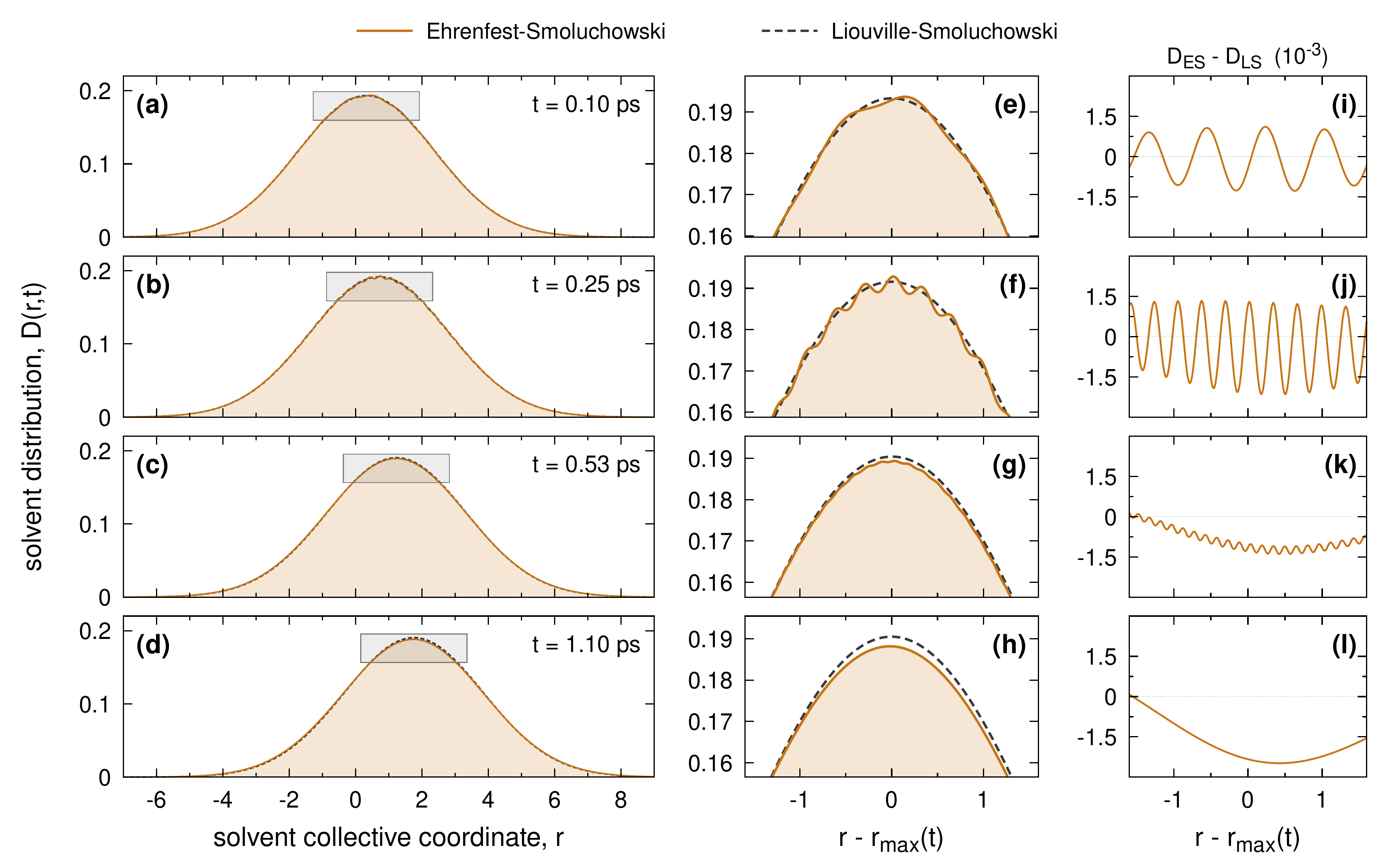}
	\caption{Solvent distribution $D(r,t)$ for DCM in CH$_2$Cl$_2$ obtained with the Ehrenfest--Smoluchowski equation (solid, shaded) and the Liouville--Smoluchowski equation (dashed), starting from the factorized initial condition with the dye in $|N\rangle$.
	Rows correspond to $t=0.10$, $0.25$, $0.53$, and $1.10\,\mathrm{ps}$.
	(a)--(d) Full distribution; the grey boxes mark the zoom windows.
	(e)--(h) Zoom on the distribution peak, located at $r_{\max}(t)=\operatorname*{arg\,max}_r D(r,t)$.
	(i)--(l) Difference between the solvent distributions obtained with the QCES and QCLS equations, $D_{\mathrm{ES}}-D_{\mathrm{LS}}$, in units of $10^{-3}$.}
	\label{fig:wiggles}
\end{figure*}

Besides the moment dynamics, the QCES model also exhibits a fine structure of the solvent distribution that is absent in the QCLS description: small, regular oscillations form on top of $D(r,t)$ at early times.
Figure~\ref{fig:wiggles} shows the classical distribution for DCM in the CH$_2$Cl$_2$ solvent as a representative example.
Within the QCES dynamics, these oscillations set in after about $0.06\,\mathrm{ps}$ and are already clearly visible at $0.10\,\mathrm{ps}$ (panels (a), (e), and (i)).
Their wavelength shortens steadily as time proceeds (panels (b), (c), (f), (g), (j), and (k)); once it approaches the grid resolution, the oscillations fade, and at $1.10\,\mathrm{ps}$ the distribution differs from the QCLS result only by a smooth offset (panels (d), (h), and (l)).
The QCLS distribution, in contrast, remains smooth at all times.
The oscillations are small, about $1\%$ of the maximum of $D(r,t)$, and therefore leave the moments in Fig.~\ref{fig:moments} essentially unaffected; they are nonetheless a robust feature of the model.
They appear in all three solvents, during $0.05$--$0.75\,\mathrm{ps}$ in DMSO and $0.09$--$1.28\,\mathrm{ps}$ in CHCl$_3$, and equally in the excited-state dynamics underlying the emission spectra of Section~\ref{subsec:Emission Spectra}, where the dye is prepared in $|Z\rangle$ on the stationary ground-state solvent distribution and oscillations are present during the first $0.5$--$0.8\,\mathrm{ps}$.

The origin of the oscillations lies in the competition of scales between the unitary solute dynamics and the solvent relaxation.
At early times, before relaxation takes over, the two systems interact while undergoing very different types of evolution: while the quantum solute undergoes Hamiltonian electronic transitions, the solvent relaxation is transiently disrupted by the interaction with the solute which induces fine-structure dynamics.
In the long term, however, the free energy must decay so that relaxation drives the overall system, thereby smoothing all the small-scale dynamics.
This transient competition between the phenomenology at different scales is intrinsic to the QCES model, which, as discussed above, applies irreversible relaxation selectively to the solvent.

Since only the QCES equation relaxes to a stationary state, as shown by both the energy components and the moments of the solvent distribution, we use these stationary states as the equilibrium initial conditions for the calculation of optical spectra in the remainder of this work.
The QCLS model is not considered further, since it does not provide the stationary reference state required for the spectroscopic analysis.

\section{Absorption and emission spectra in QCES theory}\label{sec:spectra}
\subsection{Dipole-dipole correlations and absorption spectra}
To calculate the optical spectra, we first define the electric dipole moment operator of the dye.
In the diabatic basis used here, only the zwitterionic state carries a permanent dipole moment, so that:
\begin{equation}\label{eq:dipole-operator}
	\widehat\mu
	=\mu_0|Z\rangle\langle Z|,
\end{equation}
where $\mu_0=28\,\mathrm{D}$ for DCM.\cite{terenziani2015}
Upon denoting $\langle\!\langle\widehat{A}\rangle\!\rangle=\operatorname{Tr}{\int}\widehat{P}\widehat{A}\,\de r$, the absorption spectrum is obtained from the real part of the Fourier transform of the dipole-dipole correlation function:\cite{Mukamel1988,DiMaioloPainelli2020}
\begin{equation}\label{eq:absorption-correlation}
	C_{\mu\mu}^{\mathrm{abs}}(t)
	=\langle\!\langle\widehat\mu(t)\widehat\mu(0)\rangle\!\rangle
	=\operatorname{Tr}{\int}\widehat\mu\,\widehat\Omega^{\mathrm{abs}}(r,t)\,\de r,
\end{equation}
where the $\widehat\mu$\,'s after the first equality are Heisenberg operators and $\widehat\Omega^{\mathrm{abs}}(r,t)$ is the corresponding spectral generating function.
Its initial condition is obtained from the stationary ground-state distribution discussed in the previous section:
\begin{equation}\label{eq:absorption-initial-operator}
	\widehat\Omega^{\mathrm{abs}}(r,0)
	=\widehat\mu\,\widehat P_{\mathrm{eq}}(r).
\end{equation}
Although the operator $\widehat\Omega^{\mathrm{abs}}(r,0)$ is generally non-Hermitian, we evolve it according to Eq.~\eqref{eq:qces}, originally formulated for the hybrid density $\widehat P$.
In the numerical implementation, the real-valued duality pairing in Eq.~\eqref{eq:real-duality-pairing} provides the real scalar inputs required by the classical transport step, even for a non-Hermitian operator.
At the start of each classical substep, the density and mean force are evaluated as $D^{\mathrm{abs}}=\operatorname{Re}\operatorname{Tr}\widehat\Omega^{\mathrm{abs}}$ and $F_r^{\mathrm{abs}}=-\operatorname{Re}\operatorname{Tr}(\widehat\Omega^{\mathrm{abs}}\,\partial_r\widehat H)/D^{\mathrm{abs}}$.
The classical substep uses these real coefficients to transport the full operator $\widehat\Omega^{\mathrm{abs}}(r,t)$, and the quantum substep acts on it by unitary conjugation.

The spectrum is computed for $E>0$ as
\begin{equation}\label{eq:absorption-spectrum}
	S(E)
	=2\Delta t\,\operatorname{Re}\sum_{n=0}^{N}w_n\bigl[C_{\mu\mu}^{\mathrm{abs}}(t_n)-C_{\mu\mu}^{\mathrm{abs}}(T)\bigr]e^{+iEt_n/\hbar},\qquad
	t_n
	=n\Delta t,\quad
	T
	=N\Delta t,
\end{equation}
with $N=999$, $\Delta t\approx0.1\,\mathrm{fs}$, and trapezoidal weights ($w_0=w_N=1/2$, $w_n=1$ otherwise).
The sum in Eq.~\eqref{eq:absorption-spectrum} is evaluated by a zero-padded fast Fourier transform, which only refines the energy grid.
We normalize the computed spectrum $S(E)$ to its peak value.

\begin{figure*}[t!]
	\centering
	\includegraphics[width=\textwidth]{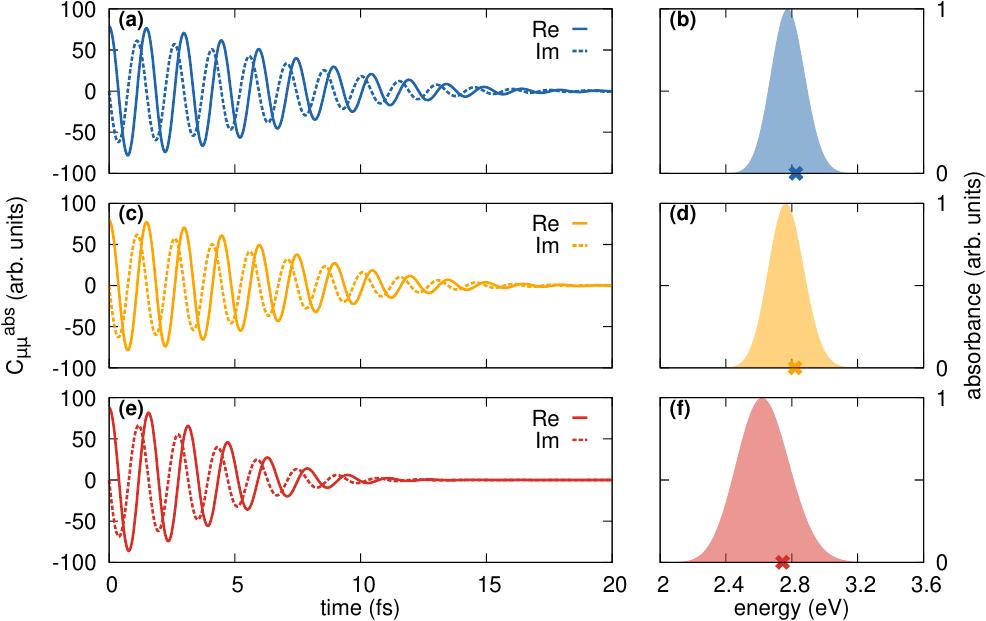}
	\caption{Calculated dipole-dipole correlation functions and absorption spectra of DCM in CHCl$_3$ (panels (a) and (b); $\tau_L=2.8\,\mathrm{ps}$, $\epsilon=0.32\,\mathrm{eV}$), CH$_2$Cl$_2$ (panels (c) and (d); $\tau_L=0.56\,\mathrm{ps}$, $\epsilon=0.35\,\mathrm{eV}$), and DMSO (panels (e) and (f); $\tau_L=2\,\mathrm{ps}$, $\epsilon=0.75\,\mathrm{eV}$).
	Panels (a), (c), and (e) show the real (solid lines) and imaginary (dashed lines) parts of $C_{\mu\mu}^{\mathrm{abs}}(t)$ as functions of time.
	Panels (b), (d), and (f) show the corresponding absorption spectra, obtained as the real part of the Fourier transform of the correlation functions.
	Crosses indicate the reference vertical transition energies obtained by exact diagonalization of the solute-solvent Hamiltonian at the ground-state equilibrium solvent coordinate for each solvent.
	Model parameters are reported in the main text.}
	\label{fig:absorption_spectra}
\end{figure*}

Figure~\ref{fig:absorption_spectra} shows the calculated absorption spectra for DCM in the three solvents discussed above.

For CHCl$_3$, the real and imaginary parts of $C_{\mu\mu}^{\mathrm{abs}}(t)$ are shown in panel (a), while the corresponding spectrum is reported in panel (b).
The spectrum has a maximum at $2.77\,\mathrm{eV}$, compared with the reference vertical transition energy of $2.83\,\mathrm{eV}$ indicated by the blue cross.
This reference energy is obtained as the difference between the excited- and ground-state potential energy curves, calculated by exact diagonalization of the solute-solvent Hamiltonian at fixed $r$ and evaluated at the ground state equilibrium solvent coordinate. 
For CH$_2$Cl$_2$ (panels (c) and (d)), the spectral maximum and the reference vertical transition energy (orange cross) are $2.76$ and $2.82\,\mathrm{eV}$, respectively.
The largest difference is found for DMSO (panels (e) and (f)), where the spectrum peaks at $2.62\,\mathrm{eV}$, whereas the reference vertical transition energy (red cross) is $2.74\,\mathrm{eV}$.

The time-domain correlations also reveal a clear solvent dependence of the spectral broadening.
Their decay occurs on similar timescales in CHCl$_3$ and CH$_2$Cl$_2$, consistent with the comparable widths of the corresponding absorption bands, whose full width at half maximum (FWHM) values are $0.24$ and $0.25\,\mathrm{eV}$, respectively.
In DMSO, instead, the dipole-dipole correlation function decays more rapidly, leading to a broader absorption band with an FWHM of $0.38\,\mathrm{eV}$.
This enhanced broadening can be traced back to the stronger solute-solvent coupling in DMSO.
As the solvent relaxation energy $\epsilon$ increases, the polar $|Z\rangle$ diabatic state is stabilized and brought closer in energy to the neutral $|N\rangle$ state.
The resulting stronger avoided-crossing interaction softens the lower adiabatic potential around its equilibrium region.
Consequently, the thermal distribution of the reaction-field coordinate becomes broader, producing a larger distribution of instantaneous transition energies.
This increased energetic disorder leads to a faster decay of the dipole-dipole correlation function and, correspondingly, to the larger spectral width observed for DMSO.

\subsection{Time-resolved emission spectra}\label{subsec:Emission Spectra}
Time-resolved emission spectra are obtained by first generating the non-equilibrium excited-state dynamics.
Photoexcitation is modeled by preparing the dye in the zwitterionic diabatic state $|Z\rangle$ while retaining the equilibrated ground-state distribution $D_{\mathrm{eq}}(r)=\operatorname{Tr}\widehat P_{\mathrm{eq}}(r)$ of the reaction-field coordinate:
\begin{equation}\label{eq:photoexcited-initial-state}
	\widehat P(r,0)
	=D_{\mathrm{eq}}(r)\,|Z\rangle\langle Z|.
\end{equation}
Because the diabatic coupling $\tau$ is finite, $|Z\rangle$ does not coincide with the upper adiabatic state.
The corresponding initial upper adiabatic state populations are $0.887$, $0.886$, and $0.871$ for CHCl$_3$, CH$_2$Cl$_2$, and DMSO, respectively, with the remaining population residing in the lower adiabatic state.
These populations remain essentially unchanged during the excited state dynamics, with variations below $0.003$.
The resulting distribution is then propagated up to a set of delay times $t_k'$.
At each delay, the instantaneous QC distribution is used to initialize the emission spectral generating function:
\begin{equation}\label{eq:emission-initial-operator}
	\widehat\Omega^{\mathrm{fl}}(r,0;t_k')
	=\widehat\mu\,\widehat P(r,t_k').
\end{equation}
The latter is propagated along the correlation time $t''$ similarly to $\widehat\Omega^{\mathrm{abs}}$.
The corresponding dipole-dipole correlation function is evaluated as:
\begin{equation}\label{eq:emission-correlation}
	C_{\mu\mu}^{\mathrm{fl}}(t'';t_k')
	=\langle\!\langle\widehat\mu(t'')\widehat\mu(0)\rangle\!\rangle_{t_k'}
	=\operatorname{Tr}{\int}\widehat\mu\,\widehat\Omega^{\mathrm{fl}}(r,t'';t_k')\,\de r.
\end{equation}
The emission spectrum at delay time $t_k'$ is obtained with the same time step, plateau subtraction, and weights as for absorption, with the correlation function truncated at $T''=N\Delta t\approx24\,\mathrm{fs}$ ($N=250$), but with the opposite Fourier sign, $e^{-iEt''_n/\hbar}$ with $E>0$.
The sign change reflects the opposite direction of the underlying transition: in absorption, the dye is promoted from the lower to the upper adiabatic state, so that the dominant contribution to $C_{\mu\mu}^{\mathrm{abs}}$ carries the phase factor $e^{-i\Delta E\,t/\hbar}$, whereas in emission it returns from the upper to the lower state, and $C_{\mu\mu}^{\mathrm{fl}}$ carries the conjugate factor $e^{+i\Delta E\,t/\hbar}$.
In both cases, the corresponding Fourier sign places the band at positive energies.

\begin{figure*}[tb]
	\begin{center}
		\includegraphics[width=\textwidth]{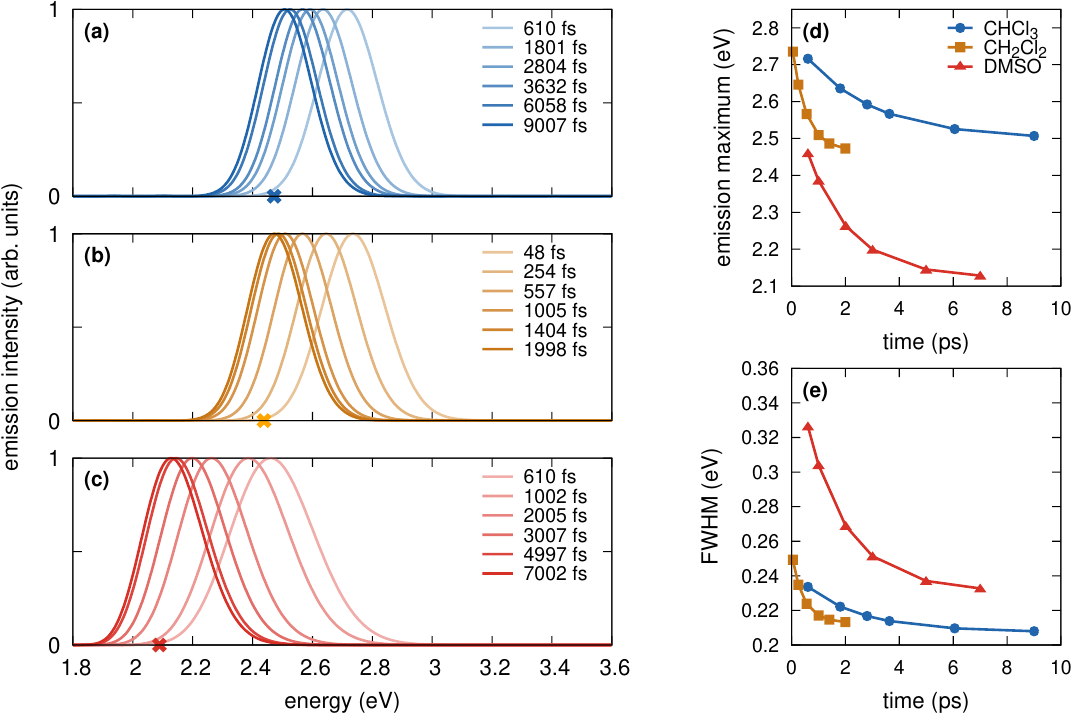}
		\caption{Time-resolved emission spectra of DCM in CHCl$_3$ (a), CH$_2$Cl$_2$ (b), and DMSO (c), evaluated at six different delay times.
		In each panel, the colored cross marks the reference vertical transition energy obtained by exact diagonalization of the solute-solvent Hamiltonian at the excited-state equilibrium solvent coordinate as relevant to the emission process.
		Panel (d) shows the time evolution of the emission maximum for the three solvents, highlighting the dynamic Stokes shift associated with solvent relaxation.
		Panel (e) reports the corresponding evolution of the full width at half maximum (FWHM) of the emission band.}
		\label{fig:emission_spectra}
	\end{center}
\end{figure*}

The resulting time-resolved emission spectra are shown in Fig.~\ref{fig:emission_spectra}.
In all three solvents, the emission band progressively shifts to lower energies as the reaction field relaxes toward the excited state equilibrium configuration, providing a direct description of the dynamic Stokes shift.
The evolution occurs on a solvent-dependent timescale of approximately $3.5\tau_L$.
The colored crosses indicate the reference vertical transition energies obtained by exact diagonalization of the solute-solvent Hamiltonian at the excited state equilibrium solvent coordinate.
For CHCl$_3$ (panel (a)), the steady-state spectrum is reached after about $9\,\mathrm{ps}$ and peaks at $2.51\,\mathrm{eV}$, compared with the reference transition energy of $2.47\,\mathrm{eV}$ (blue cross).
For CH$_2$Cl$_2$ (panel (b)), relaxation is considerably faster, with the spectrum converging after about $2\,\mathrm{ps}$ to a maximum at $2.47\,\mathrm{eV}$, compared with the reference transition energy of $2.44\,\mathrm{eV}$ (orange cross).
Finally, for DMSO (panel (c)), the steady-state spectrum is reached after approximately $7\,\mathrm{ps}$ and peaks at $2.13\,\mathrm{eV}$, whereas the reference energy from diagonalization is $2.09\,\mathrm{eV}$ (red cross).
Thus, the steady-state emission maxima obtained from the dynamical calculations lie slightly above the corresponding reference vertical transition energies, with similar differences of $0.03$--$0.04\,\mathrm{eV}$ across the three solvents.

The time dependence of the emission maximum is summarized in panel (d), which highlights the progressive red shift and the different relaxation timescales of the three solvents.
The evolution of the spectral width is reported in panel (e).
In all cases, the FWHM decreases as the solvent relaxes.
For CHCl$_3$, it decreases from $0.23\,\mathrm{eV}$ at $t_k'=610\,\mathrm{fs}$ to $0.21\,\mathrm{eV}$ at $t_k'=6058\,\mathrm{fs}$, while for CH$_2$Cl$_2$ it decreases from $0.25\,\mathrm{eV}$ at $t_k'=48\,\mathrm{fs}$ to the same limiting value of $0.21\,\mathrm{eV}$ at $t_k'=1998\,\mathrm{fs}$.
The similar final widths reflect the comparable solute-solvent coupling strengths of the two solvents, whereas the different relaxation times follow from their markedly different solvent mobilities.
In DMSO, the FWHM decreases more substantially, from $0.33\,\mathrm{eV}$ at $t_k'=610\,\mathrm{fs}$ to $0.23\,\mathrm{eV}$ at $t_k'=7002\,\mathrm{fs}$.

The spectral narrowing can be understood in terms of the same avoided crossing discussed for the absorption spectra.
Increasing the stabilization of the polar $|Z\rangle$ state brings the two diabatic states closer in energy, so that level repulsion softens the lower adiabatic surface while hardening the upper one.
During excited state relaxation, the system evolves on this stiffer upper surface, and the corresponding reaction-field distribution becomes progressively narrower.
The resulting reduction in the spread of solvent configurations decreases the distribution of instantaneous emission energies, leading to the narrowing of the time-resolved emission band.

The solvent relaxation can be further quantified through the normalized spectral response function $S_\nu(t')$, obtained from the time-dependent position of the emission maximum:\cite{Horng1995}
\begin{equation}\label{eq:spectral-response}
	S_\nu(t_k')
	=\frac{E_{\max}(t_k')-E_{\max}(\infty)}{E_{\max}(0)-E_{\max}(\infty)}.
\end{equation}
Here, $E_{\max}(t_k')$ is the emission maximum at delay time $t_k'$ and $E_{\max}(\infty)$ corresponds to the fully relaxed emission maximum.
Following the conventional definition of the spectral response function, $E_{\max}(0)$ refers to an effective zero-time condition in which the intramolecular vibrational degrees of freedom are assumed to have relaxed, while the solvent remains frozen in its initial, unrelaxed configuration.\cite{Fee1994}
Since vibrational degrees of freedom are not treated explicitly in the present model, we approximate this reference by the transition energy obtained from exact diagonalization of the solute-solvent Hamiltonian at the ground state equilibrium solvent coordinate.
With this normalization, $S_\nu$ measures the progress of solvent relaxation from the initially prepared state toward the relaxed excited state configuration.

\begin{figure*}[tb]
	\begin{center}
		\includegraphics[width=0.6\textwidth]{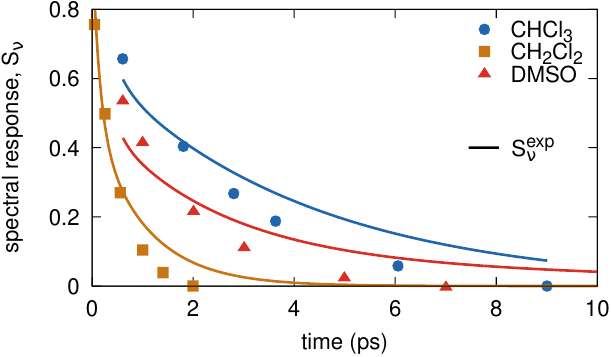}
		\caption{Normalized spectral response function, $S_\nu(t')$, as a function of time delay for CHCl$_3$ (blue circles), CH$_2$Cl$_2$ (orange squares), and DMSO (red triangles).
		Points denote calculated $S_\nu$ at discrete delay times, and solid lines denote the corresponding experimental response curves, $S_\nu^{\mathrm{exp}}$, from Ref.~\citenum{Horng1995}.}
		\label{fig:snu}
	\end{center}
\end{figure*}

Figure~\ref{fig:snu} shows $S_\nu(t_k')$ for DCM in CHCl$_3$, CH$_2$Cl$_2$, and DMSO, evaluated pointwise at the six delay times considered in Fig.~\ref{fig:emission_spectra}.
The corresponding experimental spectral response functions reported in Ref.~\citenum{Horng1995} are shown as continuous curves.
The calculated points reproduce well the characteristic relaxation trends and the markedly different timescales of the three solvents.
In particular, the faster response of CH$_2$Cl$_2$ and the slower relaxation observed in CHCl$_3$ and DMSO are captured by the model.
The agreement with experiment is notable given the compact description of the solvent through a single overdamped reaction-field coordinate, indicating that this coordinate retains the dominant features of the solvent reorganization probed by the dynamic Stokes shift.

\section{Conclusions}\label{sec:conclusions}
Following previous proposals\cite{BoEtAl11,Burghardt2006,HuEtAl12} in solvation theory, we have developed a quantum-classical DDFT framework for describing the coupled dynamics of a molecular electronic subsystem and an overdamped polar solvent represented through a reaction-field coordinate (cf. Fig.~\ref{fig:model}).
The central quantity of the formulation is an operator-valued distribution $\widehat P(r,t)$, which retains the quantum electronic degrees of freedom while resolving their correlations with the classical solvent coordinate.
Starting from a free-energy-driven construction, we derived a QCES equation in which the irreversible dynamics is associated with the classical solvent distribution and the free energy is guaranteed to be non-increasing.
This structure ensures that the stationary solution is a thermal state, so that detailed balance is retained.

We showed the importance of this structure by comparing the QCES equation with the QCLS model (cf. Fig.~\ref{fig:equilibrium}), previously derived in Ref.~\citenum{DiMaioloPainelli2020} from the quantum-classical Liouville equation.
For DCM in CHCl$_3$, CH$_2$Cl$_2$, and DMSO, the QCES dynamics drives the coupled dye-solvent system toward a stationary thermal state, whereas the QCLS model develops an unstable long-time behavior and fails to provide a suitable thermal reference state.
Using DCM in DMSO as a representative case, we showed that the mean and variance of the solvent distribution reach stationary values in QCES, whereas in QCLS the mean continues to drift and the distribution narrows at long times (cf. Fig.~\ref{fig:moments}).
This difference in the mean is also observed in the other two solvents, where its persistent drift in QCLS signals a failure to reach thermal equilibrium.
Our QCES simulations also reveal small, transient oscillations in the density $D(r,t)$ in all three solvents, both during ground state relaxation and after photoexcitation (cf. Fig.~\ref{fig:wiggles}).
These oscillations arise when solvent transport carries the spatial pattern of coherent electronic population oscillations into the solvent distribution.
This fine-structure dynamics is a distinctive feature of QCES: it reveals the influence of electronic coherence on the shape of the solvent distribution, even though the oscillations have little effect on its mean and variance.
In QCLS, diffusion of $\widehat P$ smooths out this fine structure.

Within the QCES framework, absorption and time-resolved emission spectra were calculated from dipole-dipole correlation functions, naturally accounting for the inhomogeneous broadening associated with thermal fluctuations of the solvent coordinate.
In the absorption spectra, the broadening is strongest in DMSO as a consequence of the stronger solute-solvent coupling and the corresponding softening of the lower adiabatic surface (cf. Fig.~\ref{fig:absorption_spectra}).
Following photoexcitation, the model captures both the dynamic Stokes shift associated with solvent relaxation and the concomitant evolution of the emission bandwidth (cf. Fig.~\ref{fig:emission_spectra}).
In particular, the emission bands progressively narrow as the solvent distribution becomes more localized on the harder upper adiabatic surface.
The steady-state emission maxima remain close to the reference vertical transition energies evaluated at the excited state equilibrium solvent coordinate.

The time evolution of the emission maxima also provides direct access to the normalized solvent spectral response function (cf. Fig.~\ref{fig:snu}).
Despite the compact description of the environment in terms of a single overdamped reaction-field coordinate, the calculated response reproduces well the characteristic experimental relaxation trends reported for the three solvents.
These results show that the present framework provides a unified description of thermal relaxation, solvent-induced spectral broadening, and non-equilibrium optical response within the same quantum-classical dynamical framework.

The approach provides a natural starting point for extensions to more complex molecular models and solvent descriptions.
In particular, future work will address additional molecular degrees of freedom and more structured solvent relaxation with the aim of assessing the range of validity of the Ehrenfest approximation.
In addition, we will consider further variants of the QC-DDFT framework aimed at overcoming the limitations generally affecting the Ehrenfest approach in nonadiabatic dynamics.


%
\begin{acknowledgement}
	We are grateful to Federica Agostini, Raymond Kapral, and Klaus Schleich for several enlightening discussions on this and related topics.
	The work of P.B. was funded by the Polish NCN project no. 2019/34/E/ST1/00390. 
	F.D.M. acknowledges financial support from Bando di Ateneo per la ricerca 2025 -- Azione C (DYNAMO, D53C25002680005) and Bando di Ateneo per la Ricerca 2026 -- Azione B (BRIGHT, D53C26003680005), computational resources provided by the High Performance Computing (HPC) facility at the University of Parma, and computing resources awarded by CINECA under the ISCRA initiative (IsCd8\_BRIGHTER).
	I.B. and C.T. acknowledge financial support by the Leverhulme Trust Research Project Grant RPG-2023-078.
\end{acknowledgement}
%


\begingroup
\linespread{1.6}\selectfont 
\bibliography{biblio}
\endgroup
\end{document}